\documentclass[structabstract]{aa}  
\usepackage{graphicx}
\usepackage{txfonts}
\usepackage{lscape}
\usepackage{longtable,lscape}
\usepackage{natbib}
\bibpunct{(}{)}{;}{a}{}{,} 
\begin{document}
   \title{The X-ray spectral properties of the AGN population in the {\it XMM-Newton} bright serendipitous survey}


   \author{A. Corral
          \inst{1}
          \and
          R. Della Ceca\inst{1}
          \and
          A. Caccianiga\inst{1}
          \and
          P. Severgnini\inst{1}
          \and
          H. Brunner\inst{2}
	  \and
          F.J. Carrera\inst{3}
          \and
          M.J. Page\inst{4}
	  \and 
          A.D. Schwope\inst{5}
          }

   \institute{INAF - Osservatorio Astronomico di Brera, via Brera 28, 20121, Milan, Italy\\
              \email{amalia.corral@brera.inaf.it}
\and Max-Planck-Institut f\"ur extraterrestrische Physik, 85478 Garching, Germany 
\and Instituto de F\'\i{}sica de Cantabria (CSIC-UC), 39005 Santander, Spain
\and Mullard Space Science Laboratory, University College London, Holmbury St Mary, Surrey RH5 6NT, UK
\and Astrophysikalisches Institut Potsdam, An der Sternwarte 16, 14482 Potsdam, Germany
}
   \date{Received , ; accepted , }
 %

 %
 
  \abstract 
{X-ray surveys are a key instrument in the study of active  galactic nuclei (AGN). Thanks to their penetrating ability, X-rays are able to map the innermost regions close to the central super massive black hole (SMBH) as well as to detect and characterize its emission up to high redshift.}  
{We present here a detailed X-ray spectral analysis of the AGN belonging to the XMM-Newton bright survey (XBS). The XBS is composed of two flux-limited samples selected in the complementary 0.5-4.5 and 4.5-7.5 keV energy bands and comprising more than 300 AGN up to redshift $\sim$ 2.4.}  
{We performed an X-ray analysis following two different approaches: by analyzing individually each AGN X-ray spectrum and by constructing average spectra for different AGN types.}  
{From the individual analysis, we find that there seems to be an anti correlation between the spectral index and the sources' hard X-ray luminosity, such that the average photon index for the higher luminosity sources ($>$ 10$^{44}$ erg s$^{-1}$) is significantly ($>$ 2$\sigma$) flatter than the average for the lower luminosity sources. We also find that the intrinsic column density distribution agrees with AGN unified schemes, although a number of exceptions are found (3\% of the whole sample), which are much more common among optically classified type 2 AGN. We also find that the so-called ``soft-excess'', apart from the intrinsic absorption, constitutes the principal deviation from a power-law shape in AGN X-ray spectra and it clearly displays different characteristics, and likely a different origin, for unabsorbed and absorbed AGN. Regarding the shape of the average spectra, we find that it is best reproduced by a combination of an unabsorbed (absorbed) power law, a narrow Fe K$\alpha$ emission line and a small (large) amount of reflection for unabsorbed (absorbed) sources. We do not significantly detect any relativistic contribution to the line emission and we compute an upper limit for its equivalent width (EW) of 230 eV at the 3$\sigma$ confidence level. Finally, by dividing the type 1 AGN sample into high- and low-luminosity sources, we marginally detect a decrease in the narrow Fe K$\alpha$ line EW and in the amount of reflection as the luminosity increases, the ``so-called'' Iwasawa-Taniguchi effect.}
{}
   \maketitle
%

\section{Introduction}
Recent deep X-rays surveys carried out by {\it XMM-Newton} and Chandra
have resolved most of the cosmic X-ray background (CXB) into discrete
sources up to energies $\sim$ 10 keV (although the resolved fraction
decreases with energy; \citealt{wor05,hick06}). The large majority of
the sources that compose the CXB are active galactic nuclei (AGN), and
CXB synthesis models make use of template AGN spectra to reproduce its
shape following the AGN unified model \citep{anto93}. The unified
model, in its simplest version, states that the differences between
the different observed AGN types are due to an orientation effect,
i.e., as the inclination angle to the observer increases, the torus
surrounding the central super massive black hole (SMBH) intercepts
more nuclear emission. The CXB is then reproduced by a mixture of AGN
spectra with different amounts of absorption.

However, there are still many unresolved questions regarding our
knowledge about AGN. For example, the predicted fraction of heavily
absorbed AGN (Compton-thick AGN) obtained from CXB synthesis models
can vary from 30\% to 9\% between different works
\citep{gilli07,tre09}. Besides, a small number of AGN that seem not to
follow the unified scheme are usually found in X-rays surveys, i. e.,
their optical characteristics do not match their observed X-ray
properties (\citealt{panessa02}, \citealt{akylas04},
\citealt{caccianiga04}, \citealt{cappi06}, \citealt{mateos05a},
\citealt{mateos05b}, \citealt{mateos10}). The evolution of these
properties through cosmic time and the possible correlation between
X-ray emission and source properties, like the bolometric luminosity
or SMBH mass, are also a matter of debate.

The frequency and properties of some individual characteristics are
also still unknown, such as the Fe K$\alpha$ emission line. This
emission line is the most commonly observed line in AGN X-ray spectra,
but its detailed study is strongly limited by the data quality and
therefore, to sources in the local Universe \citep{nandra07}. To study
its characteristics up to high redshifts, X-ray spectra have to be
stacked to improve the signal-to-noise ratio (SNR)
\citep{corral08,streb,brusa}. Another intriguing AGN feature is the
soft-excess emission in type 1 AGN, whose origin is still
unclear. Possible suggested explanations go from continuum emission
\citep{ross92,shimura93,kawa01} to atomic processes
\citep{crummy06,midd07}.

Deep and medium surveys often lack good quality X-ray and
multi-wavelength data, which limits the results, while samples
composed by high-quality data are usually not well-defined
flux-limited samples, which limits the applicability of the
results. Given all that, well-defined X-ray samples, that contain both
a significant number of reliable identifications and good enough X-ray
data quality are the key to test the current hypotheses and to link
the nearby and distant universe. We present here a detailed X-ray
spectral analysis of the AGN within the XBS sample, which is composed
of two flux-limited samples that are almost completely identified
(identification rate $\sim$ 95\%) and containing more than 300
AGN. Given the availability of both reliable optical spectroscopic
identifications and good quality X-ray spectral data, this sample is
the perfect laboratory to test AGN models and to better constrain the
AGN properties and their evolution.

We assume the cosmological model H$_{O}$=65 km s$^{-1}$ Mpc$^{-1}$,
$\Omega_{\lambda}$= 0.7 and $\Omega_{M}$=0.3 throughout this
paper. Reported errors are at 90\% confidence level unless stated
otherwise.

\section{The XBS AGN sample}
The sample of 305 AGN discussed here (XBS AGN sample hereafter) has
been extracted from the {\it XMM-Newton} bright serendipitous
survey\footnote{The XMM-Newton Bright Serendipitous Survey is one of
  the research programs conducted by the XMM-Newton Survey Science
  Center (SSC, see http://xmmssc-www.star.le.ac.uk.) a consortium of
  10 international institutions, appointed by ESA to help the SOC in
  developing the software analysis system, to pipeline process all the
  XMM-Newton data, and to exploit the XMM serendipitous
  detections. The {\it Osservatorio Astronomico di Brera} is one of
  the Consortium Institutes.}.

The XBS consists of two flux-limited serendipitous (i.e. the targets
of the XMM-Newton pointings were excluded) samples of X-ray selected
sources at high galactic latitude ($|b| >20^o$): the XMM bright
serendipitous survey sample (BSS, 389 sources) and the XMM hard bright
serendipitous survey sample (HBSS, 67 sources, with 56 sources in
common with the BSS sample) having an EPIC MOS2 count rate limit,
corrected for vignetting, of $10^{-2}$ cts/s and 2 $\times$ 10$^{-3}$
cts/s in the 0.5-4.5 keV and 4.5-7.5 keV energy bands, respectively;
the flux limit is $\sim$ 7 $\times$ 10$^{-14}$ erg cm$^{-2}$ s$^{-1}$
in both energy selection bands.

The details on the {\it XMM-Newton} fields selection strategy and the
source selection criteria of the XMM BSS and HBSS samples are
discussed in \citet{dellaceca04}, while a description of the optical
data and analysis, of the optical classification scheme and the
optical properties of the extragalactic sources identified so far is
presented in \citet{caccianiga07,caccianiga08}. The optical and X-ray
properties of the Galactic population are discussed in
\citet{lopez07}. Previous X-rays spectral analyses of parts of the XBS
sample have already been reported in previous works. In
\citet{caccianiga04}, the X-ray spectral analysis of a subsample
extracted form the HBSS sample is reported. \citet{galbiati05}
performed an analysis of the radio-loud AGN within the
XBS. \citet{severgnini03} unveiled the AGN-nature of three sources
previously considered as normal galaxies. Finally, \citet{dellaceca08}
presented the cosmological properties of the HBSS AGN sample.

\subsection{AGN classification}
The current classification breakdown of the XBS sample, which relies
on dedicated optical spectroscopy, is as follows: 305 AGN (including 5
BL Lacs), 8 clusters of galaxies\footnote{The sample of cluster of
  galaxies is neither statistically complete nor representative of the
  cluster population because the source detection algorithm used in
  the construction of the sample is optimized for point-like
  sources.}, 2 normal galaxies and 58 X-ray emitting stars, see
Table~\ref{classification} for a detailed summary. For 25 out of the
305 AGN that composed our sample, redshift and classification are
reported here for the first time. These new identifications are marked
in boldface in Table~\ref{splfit} (columns 2 and 3). The XBS AGN
sample contains 35 sources that are optically classified as elusive
AGN, i.e., sources for which a classification cannot be derived solely
from our optical spectroscopy, although the redshift can be
measured. These are sources that are characterized by a
significant/dominant contamination of star-light from the host galaxy
in the optical spectrum \citep{caccianiga07}. Even if the presence of
an AGN in these sources is somehow suggested by the detection of a
broad or strong emission line, the ``dilution'' caused by the host
galaxy is critical because it avoids the quantification of the optical
absorption that is necessary to classify a source as a type 1 or type
2 AGN. For these sources, the type 1/2 classification is assigned as a
function of the absence/presence of a significant amount of intrinsic
absorption in their X-ray spectra. There is one case however,
XBSJ012654.3+191246, in which a type 1/2 classification cannot be
inferred from either the optical or the X-ray data, and accordingly
this source is classified as an AGN of uncertain type. At the time of
writing, 27 X-ray sources belonging to the BSS sample are still
unidentified. Out of these 27, two also belong to the HBSS sample,
which results in a level of identification of 93\% and 97\% for the
BSS and the HBSS samples, respectively.

\begin{table*}
\caption{XBS classification summary.}
\label{classification}      
\centering
\begin{tabular}{l c c c c c c c c}   
\hline\hline       
Sample & Type 1 AGN & Type 2 AGN & BL Lacs & Stars & Clusters & Galaxies & Unidentified & AGN uncertain type\\
\hline
\hline
BSS & 269(41) & 19(10) & 5 & 58(2) & 8(1) & 2 & 27(2) & 1\\
HBSS & 42 & 20 & \dots & 2 & 1 & \dots & 2 & \dots\\
\hline
Total & 270 & 29 & 5 & 58 & 8 & 2 & 27 & 1 \\      
\hline
\hline
\end{tabular}
\begin{list}{}{}
\item The numbers between parenthesis for the BSS sample correspond to the number of sources in common with the HBSS sample.
\end{list}
\end{table*}

\addtocounter{table}{1} 
\addtocounter{table}{1}

\subsection{X-ray data}

The XBS source sample was defined using only the data from the MOS2
detector. However, to increase the statistics, the data from the MOS1
and the pn detectors were considered when available and were used for
our spectral analysis.

In Table~\ref{obsids} we report the data used for the X-ray spectral
analysis of each source: Source name; {\it XMM-Newton} observation ID;
XMM-Newton filter for each detector; the values of Galactic column
densities toward the used {\it XMM-Newton} pointings; resulting
exposure time for each detector after removing high-background
intervals; total counts for all available detectors in the 0.3-10 keV
band, and corresponding sample. To increase the number of counts for
the lowest quality spectra, we searched the {\it XMM-Newton} archive
for additional observations and selected those with the longest MOS2
exposure times; we preferred not to combine different observation data
sets to minimize possible problems related to source variability. As a
result, some of the data sets used in this analysis are different from
the ones used in \citet{dellaceca04} for the definition of the sample.

\begin{figure*}[ht]
\begin{center}
$$
\begin{array}{cc}
\includegraphics[angle=0,width=9cm]{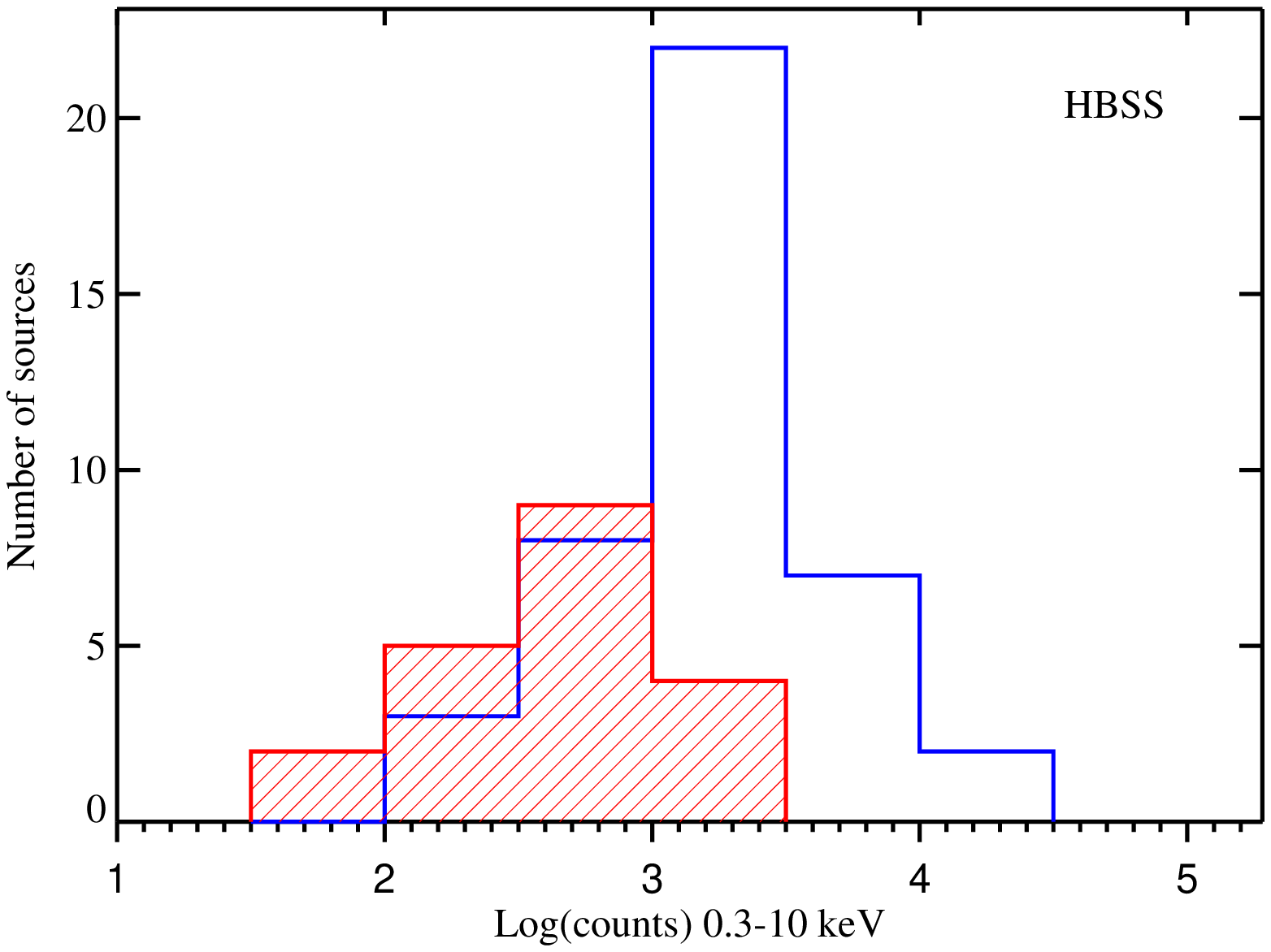} & \includegraphics[angle=0,width=9cm]{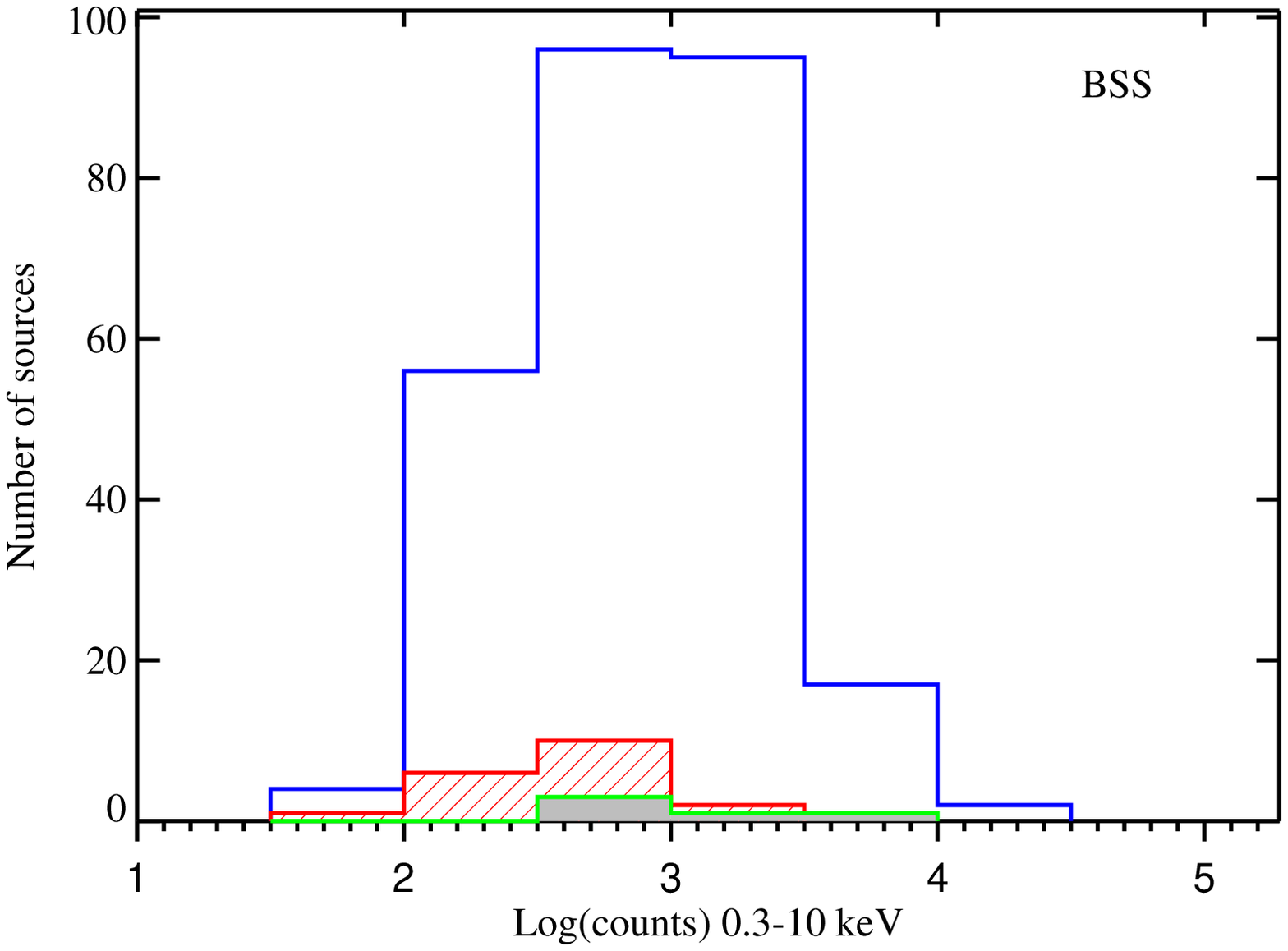}\\
\includegraphics[angle=-90,width=9cm]{15227fg1c.ps} & \includegraphics[angle=-90,width=9cm]{15227fg1d.ps}\\
\end{array}
$$
\end{center}
\caption{Top panels: Counts distribution for the HBSS (left) and the
  BSS (right) samples. Empty histograms correspond to type 1 AGN and
  dashed histograms to type 2 AGN. Filled histogram for the BSS
  corresponds to BL Lacs. Bottom panels: X-ray intrinsic luminosity
  versus redshift for the HBSS (left) and the BSS (right). Squares and
  circles correspond to sources classified as type 1 and type 2 AGN,
  respectively. Triangles on the BSS sample represent BL Lacs.}
\label{counts}
\end{figure*}

The {\it XMM-Newton} data were cleaned and processed with the {\it
  XMM-Newton} Science Analysis Software (SAS) and were analyzed with
standard software packages ({\tt Ftools}; {\tt Xspec},
\citealt{arnaud}). Event files produced from the pipeline were
filtered from high-background time intervals and only events
corresponding to pattern 0-12 for MOS and 0-4 for pn were used. All
spectra were accumulated from a circular extraction region with a
radius of $\sim 20\arcsec$-30$\arcsec$, depending on the source
off-axis distance. Background counts were accumulated in nearby
circular source free regions which an area usually about a factor
$\sim$4 larger than the one used to extract the source counts. To
improve the statistics, the MOS1 and MOS2 spectra obtained with the
same filters were combined {\it a posteriori} by using the FTOOLS task
{\it mathpha}. The X-ray spectra usually cover the 0.3$-$10 keV energy
range; the total (MOS1+MOS2+pn) counts range from $\sim$ 100 to $\sim$
10$^{4}$, as can be seen in Fig.~\ref{counts}.

The ancillary response matrix and the detector response matrix were
created by the XMM-SAS task {\it arfgen} and {\it rmfgen} at each
source position in the EPIC detectors. For the MOS1 and MOS2
detectors, and provided that the observations were carried out by
using the same filter, ancillary and detector response matrices for
each source were combined by using {\it addrmf} and {\it addarf}.

\section{X-ray spectral analysis}
The availability of good {\it XMM-Newton} data for the sources in the
XBS sample, which spans the energy range between $\sim 0.3$ and $\sim
10$ keV, allow us to perform a reliable X-ray spectral analysis for
almost every AGN studied here. For 111 AGN an X-ray spectral analysis
was already reported and discussed in \citet{severgnini03},
\citet{caccianiga04,caccianiga07}, \citet{galbiati05} or in
\citet{dellaceca08}; for the remaining AGN the main X-ray spectral
properties and parameters are discussed in detail here for the first
time. Note, however, that there could be small differences in the
best-fit model and parameters already published and the ones presented
here owing to the different {\it XMM-Newton} observations used and/or
our different way of defining the best-fit model for each source.

We grouped the spectra in bins containing at least 10 to 30 (depending
on the spectral quality) source+background counts to use the
$\chi^{2}$ minimization technique. We fitted pn and MOS spectra
simultaneously in the 0.3-10 keV band with {\tt Xspec} version
12.5.0. We tied together all pn and MOS parameters except for a
relative normalization, which accounts for the differences between pn
and MOS flux calibrations. In the following, derived fluxes and
luminosities refer to the MOS2 calibration.

To ensure a spectral analysis as uniform as possible, we defined a
threshold of 10\% for the null hypothesis probability to distinguish
between an acceptable and an unacceptable fit, i.e., we consider as
our best-fit model the simplest model for which the probability is
$>$10\%.

As a starting point for the spectral modeling we first considered a
simple absorbed power-law model that takes into account both the
Galactic hydrogen column density along the line of sight (from
\citealt{dick90}) and a possible intrinsic absorption at the source
redshift ({\tt Xspec} model: {\tt wabs*zwabs*zpo}). In the X-ray
spectral modeling we made use of the redshifts obtained from the
optical spectroscopy.

The results for this simple fit are shown in Table~\ref{splfit} along
with the corresponding Galactic de-absorbed flux and intrinsic
luminosity in the standard hard (2-10 keV) energy band. In some cases,
the spectral quality does not allow us to constrain the power-law
photon index ($\Gamma$) and the intrinsic absorption at the same
time. In other cases, the resulting photon index is $\sim$ 1, much
lower than the typical values for unabsorbed AGN\footnote{ An
  alternative possibility is that these sources with an observed flat
  spectrum are Compton-thick AGN (i.e. sources with N$_H>$10$^{24}$
  cm$^{-2}$) in which all the direct emission is suppressed and only
  reflected emission is observed (in the 2-10 keV band). This effect,
  combined with the low statistics, may mimic a flat
  spectrum. However, this hypothesis does not seem to be valid in our
  sources because in all cases we find a significant amount of
  absorption (but not in the Compton-thick regime) even when leaving
  the photon index as a free parameter (N$_H$ from
  3.5$\times$10$^{21}$ to 2$\times$10$^{23}$ cm$^{-2}$).  We conclude
  that the best explanation for the sources with a very flat spectral
  index is the combination of (mild) absorption and the low
  statistics}. In those cases, we fixed $\Gamma$ to 1.9, a common
value for unabsorbed AGN
\citep{caccianiga04,mateos05a,mateos05b,galbiati05,tozzi06,mateos10}. If
there was no intrinsic absorption detected, an upper limit, at 90\%
confidence level, is given. The simple absorbed power-law model gives
a good fit for 263 sources, but seems to fail in reproducing the
spectral shape for 41 sources, marked with a $^{p}$ in
Table~\ref{splfit}. All X-ray spectra corresponding to sources
classified as BL Lacs are well fitted by the simple power-law model.

For the 41 sources that are not well fitted by an absorbed power law,
we tried several additional components to the absorbed power law
model. We accept any of these additional components if the improvement
of the fit was larger than 95\% as measured by an F-test. These
additional components are

\begin{itemize}
\item {\bf Leaky absorbed power-law:} An additional unabsorbed
  power-law component, with the same photon index as the direct one,
  representing scattered emission into our line of sight ({\tt Xspec}
  model: {\tt wabs(zwabs*zpo+zpo)}). This model can also account for
  partially covered emission.
\item {\bf Ionized absorption:} ({\tt Xspec} model: {\tt
  wabs*zwabs*absori*zpo}, \citealt{mag}), since signatures of
  absorption from partially ionized gas have been found to be a quite
  common characteristic in the spectra of Seyfert galaxies.
\item {\bf Reflected component:} to account for a spectral hardening
  or change of curvature at high energies because of Compton
  reflection from neutral material ({\tt Xspec} model: {\tt
    wabs(zwabs*zpo+pexrav)}, \citealt{mag}). We fixed the inclination
  angle to $\sim$ 60deg, an average value for Seyfert galaxies,
  because the spectral quality does not allow us to constrain it and
  the reflection factor R (R=$\Omega/2\pi$) at the same time.
\item {\bf Thermal component} to account for soft emission lines that
  could arise from ionized material far from the central source, like
  the narrow-line region (NLR) ({\tt Xspec} model: {\tt
    wabs(zwabs*zpo+mekal)}, \citealt{mewe,lied}). Although the NLR is
  likely photoionized, we can approximate the resulting spectral shape
  by using this collisional model given the spectral resolution for
  the EPIC cameras.
\item {\bf Lines and edges} ({\tt zgauss, zegde}) to model emission lines
  (such as the Fe K$\alpha$ emission line, the most commonly observed
  one in AGN X-ray spectra) and absorption edges. Energies were left
  free to vary.
\item {\bf A phenomenological black body model} ({\tt Xspec} model:
  {\tt wabs(zwabs*zpo+zbb)} to account for featureless soft-excess
  emission.
\end{itemize}
If different additional components significantly improved the fit, we
selected the model that was more physically plausible and/or gave
better residuals. These cases, 28 AGN, are discussed in more detail in
the appendix. There are also two cases in which more than one
additional component are required to obtain an acceptable fit: in both
cases, one of the required additional components turned out to be an
emission line. A summary of the models required during the spectral
fit is shown in Table~\ref{specfit}, while the results for the
additional component fits are presented in Table \ref{bbfit} to
Table~\ref{pledgefit}. All models that significantly improve the
simple absorbed power-law fit are shown for each source in Tables
\ref{bbfit} to \ref{pledgefit}. The model we considered as the
``best-fit'' for each source is marked in boldface in those tables and
its parameters are the ones we consider in the interpretation of the
results.
\begin{figure*}[ht]
  \centering
$$
  \begin{array}{cc}
    \includegraphics[angle=0,width=9cm]{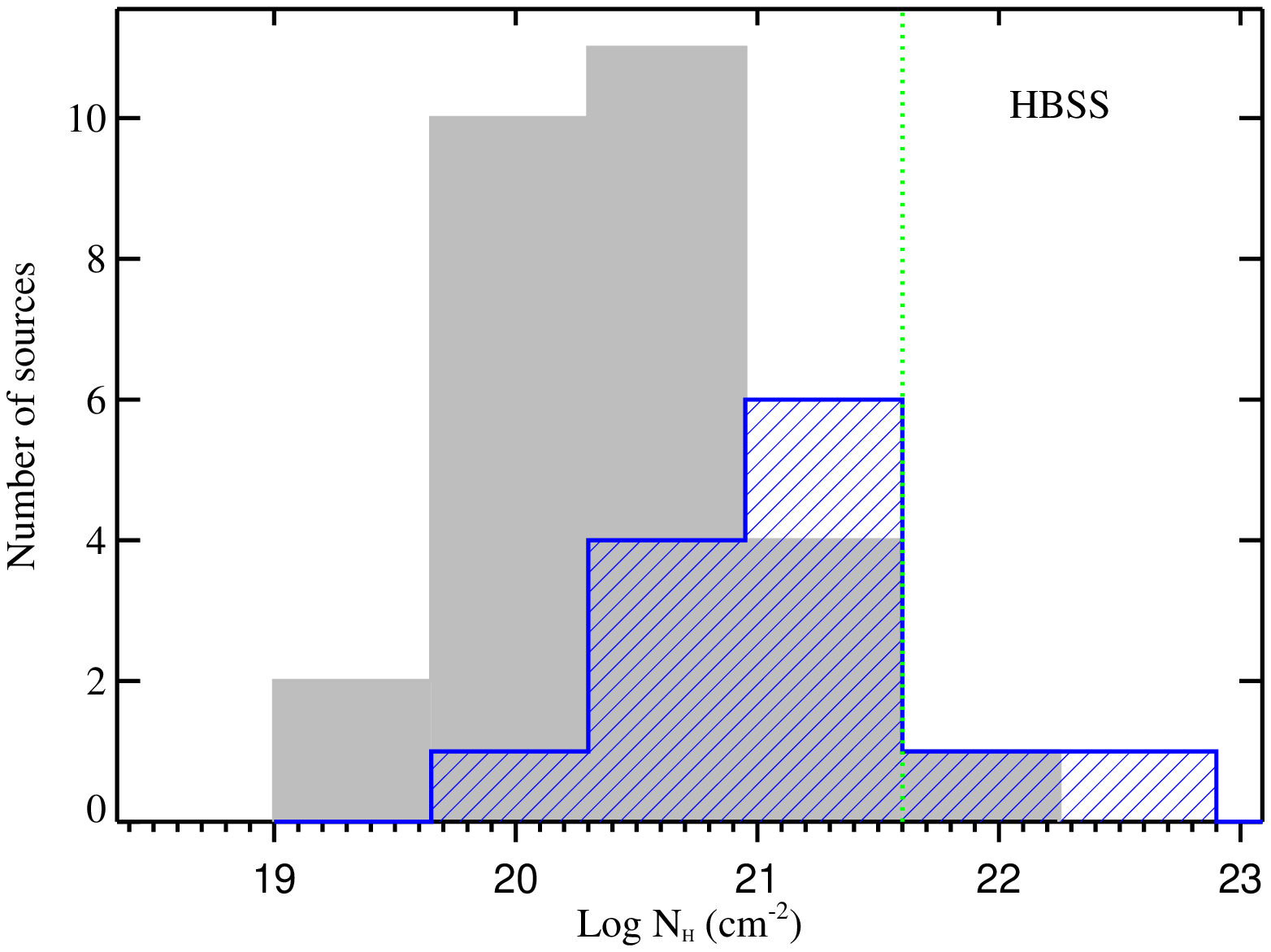} & \includegraphics[angle=0,width=9cm]{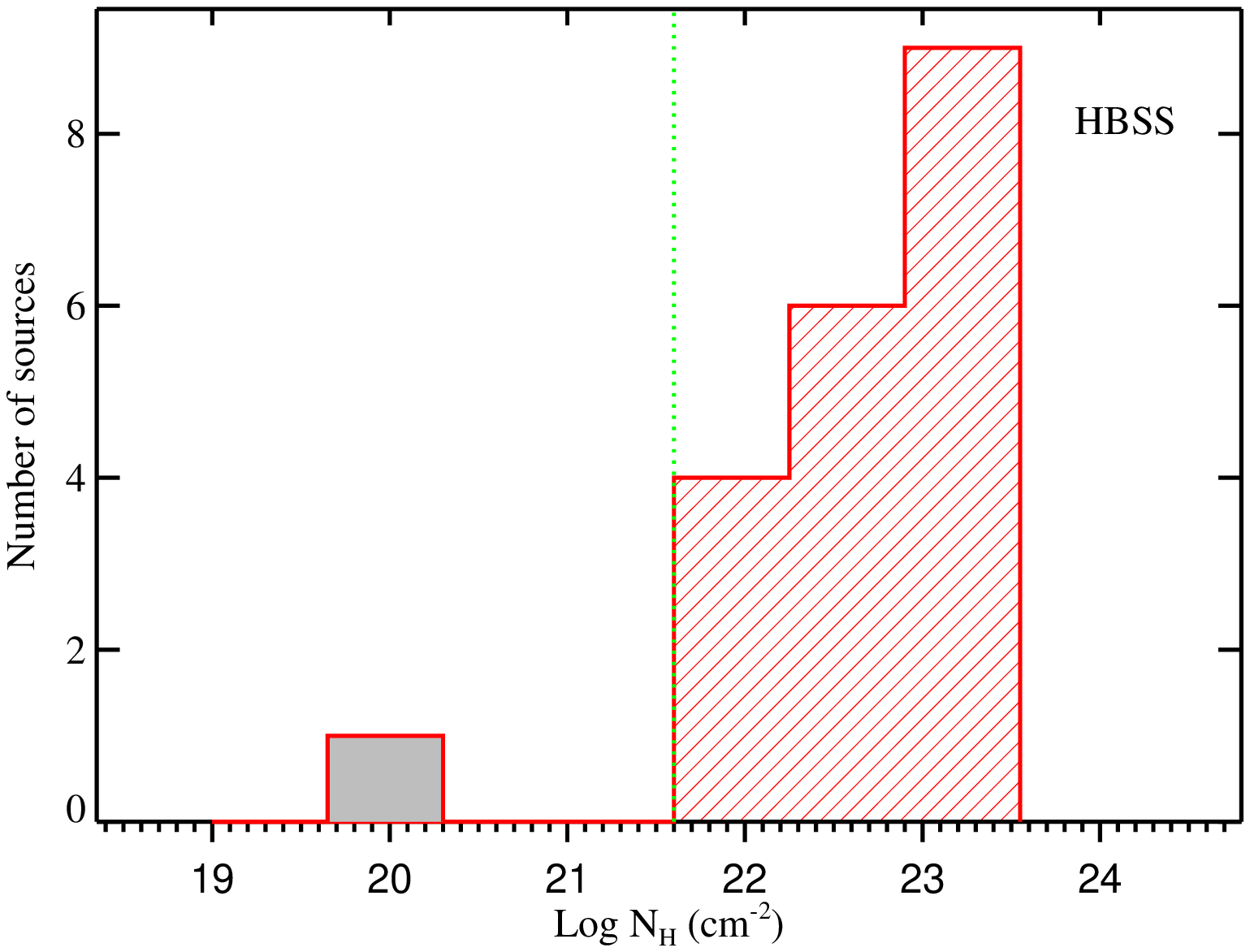}\\
    \includegraphics[angle=0,width=9cm]{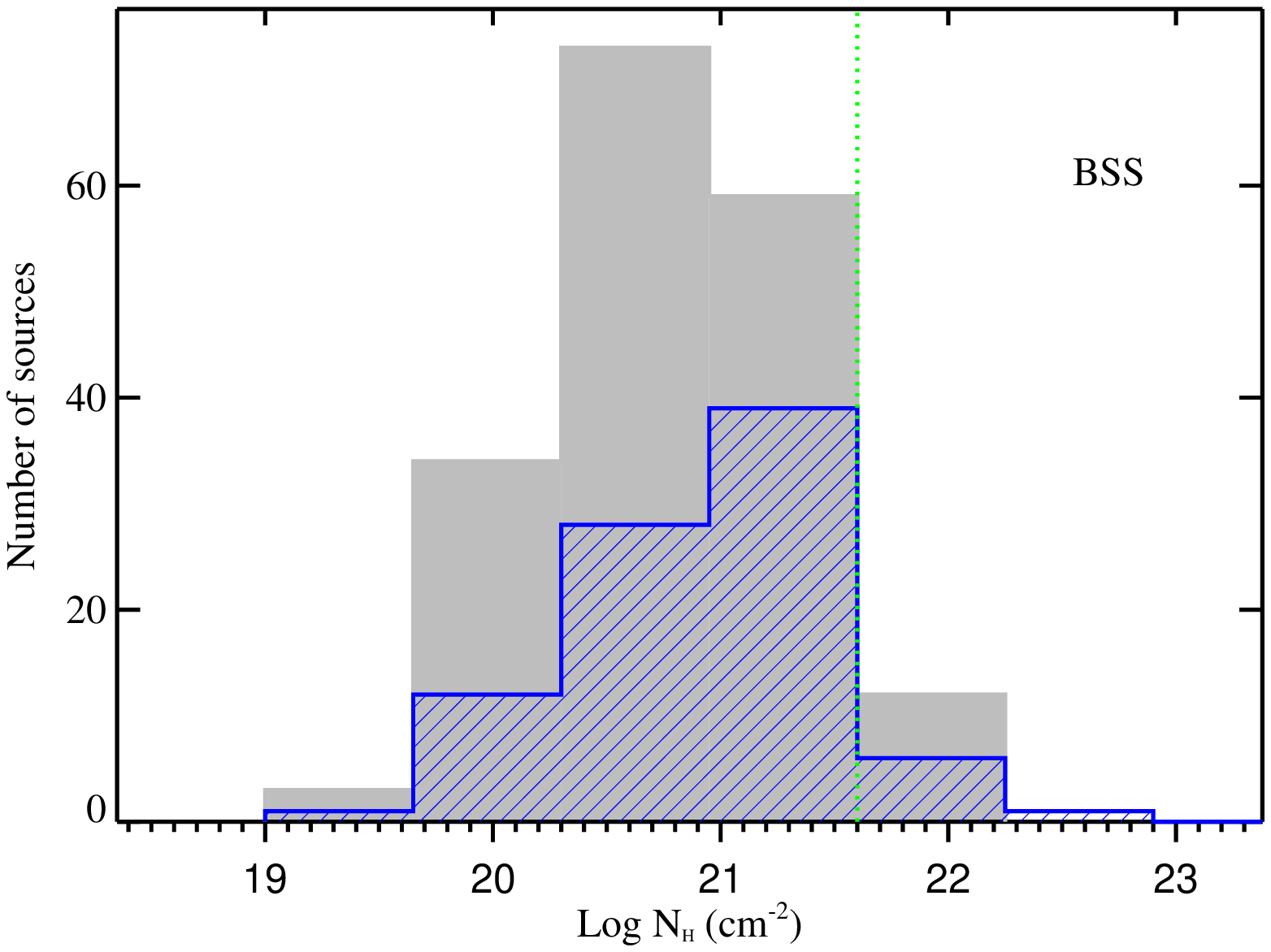} & \includegraphics[angle=0,width=9cm]{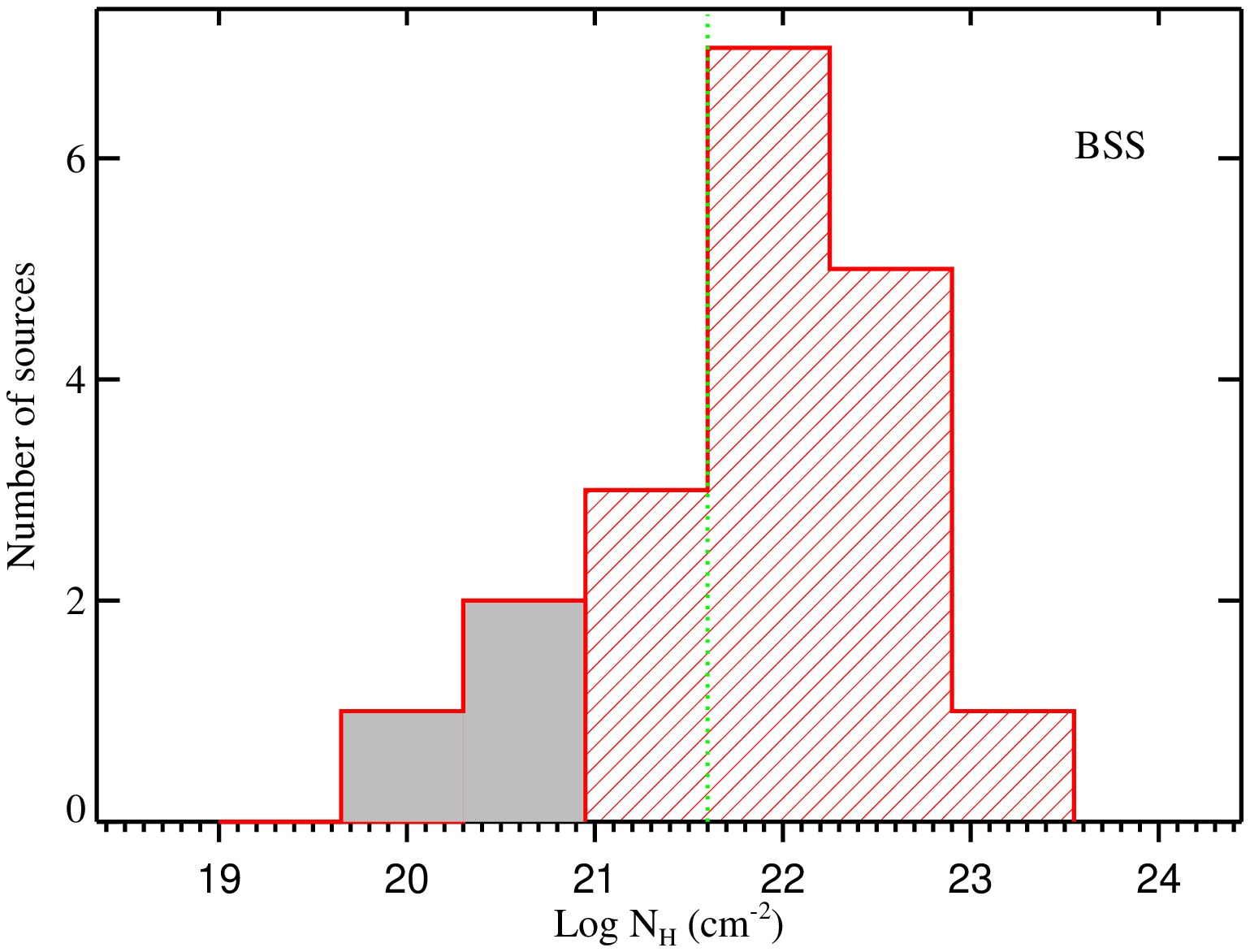}\\
  \end{array}
$$ 
 \caption{Intrinsic absorption distribution corresponding to the HBSS
   (top) and BSS (bottom). Left and right panels correspond to type 1
   and type 2 AGN, respectively. Dashed histograms correspond to
   detected intrinsic absorption, whereas filled histograms correspond
   to upper limits. Vertical dotted lines mark the threshold used to
   distinguish between absorbed and unabsorbed sources following the
   prescription of \citet{caccianiga07}.}
  \label{nhdist}
\end{figure*}

\begin{table}
\caption{X-ray spectral fit}
\label{specfit}      
\centering
\begin{tabular}{l c c c}   
\hline\hline       
Model & Type 1 AGN & Type 2 AGN & BL Lacs\\
\hline
SPL & 243 & 24 & 5\\
Leaky & 1 & 3 & \dots \\
PL+T & \dots & 2 & \dots \\
PL+BB & 13 & \dots & \dots  \\
WAPL & 5 & \dots & \dots \\
PL+R & 4 & \dots & \dots \\
PL+E & 3 & \dots& \dots \\
Leaky+L & 1 & 1 & \dots \\
\hline
\hline
\end{tabular}
\begin{list}{}{}
\item SPL: absorbed power law; Leaky: absorbed plus unabsorbed power
  law. PL+T: absorbed power law plus thermal component; PL+BB:
  absorbed power law plus black body component; WAPL: ionized absorbed
  power law; PL+R: absorbed power law plus neutral reflection
  component; PL+E: absorbed power law plus absorption edge; Leaky+L:
  leaky model plus emission line.
\end{list}
\end{table}

In 11 cases still no acceptable fit was found. This can be simply due
to a statistical effect given the 10$\%$ probability limit imposed to
consider a fit as acceptable. Roughly, 10\% of the sources that
actually display a power-law shape are expected to be not acceptably
fitted by this model. We did not find these 11 sources to share any
common spectral characteristic. For 9 of them, we found that no
additional component significantly improved the fit, and we
accordingly assume the simple absorbed power-law fit as our best fit
and used the data in Table~\ref{splfit} in the subsequent
analyses. For the remaining two sources, XBSJ021822.2--050615 and
XBSJ153456.1+013033, a leaky absorber plus an emission line and a
black body, respectively, did improve the fit significantly,
consequently we consider these models as our best-fit model although
the probability is still $<$10\%.

\section{Intrinsic absorption}

We measured intrinsic absorption in excess of the Galactic one for 119
sources, 88 type 1 and 27 type 2 AGN. This absorption is significant
(F-test $>$ 95\%) in 56 cases (30 type 1 and 26 type 2 AGN). For the
remaining 65 sources for which the significance is below 95\%, the
measured amount of absorption is very low, except for one case,
XBSJ161820.7+124116. The low significance in this case is likely owing
to the extremely low number of counts in the available spectrum. The
intrinsic absorption distribution for both studied samples is shown in
Fig.~\ref{nhdist}.

In 17 cases (3 type 1 and 14 type 2 AGN) and because of the poor
statistics, we fixed the value of $\Gamma$ to 1.9 (which corresponds
to the average value found for unabsorbed AGN) to better constrain the
intrinsic absorption. For 7 out of these 17 AGN, $\Gamma$ could be
determined (although with large errors), but turned out to be flatter
than the flattest $\Gamma$ found for unabsorbed AGN ($\Gamma \sim
1.5$). This is probably owing to the low statistics available for the
spectral analysis, which do not allow us to adequately constrain at the
same time both the spectral index and the intrinsic absorption in
these cases. We note here that the best fit N$_H$ obtained with free
$\Gamma$ is usually within the reported errors obtained when fixing
$\Gamma=1.9$; furthermore the variation of N$_H$ is such that this
problem does not have any effect on the X-ray source classification
used here (absorbed vs. unabsorbed) or on the N$_H$ distribution.

\subsection{{\bf X-rays versus optical absorption}}
Following the criteria described in \citet{caccianiga08}\footnote{The
  optimum dividing line between optical type 1/2 classification is
  found to correspond to an optical extinction of A$_{V}$ $\sim$ 2
  mag, which, assuming a Galactic A$_{V}$/N$_{H}$ ratio, implies a
  column density of N$_{H}$ $\sim$ 4$\times$10$^{21}$ cm$^{-2}$ in
  X-rays.}, we defined a sourced as absorbed if the measured intrinsic
column density is higher than 4$\times$10$^{21}$ cm$^{-2}$. According
to this criterion, we find that 31 sources are absorbed, 8 type 1 (3\%
of type 1 AGN) and 23 type 2 AGN (80\% of type 2 AGN). We find that
the fraction of absorbed sources and the amount of absorption is much
higher for type 2 AGN than for type 1 AGN, in agreement with the AGN
unified picture. However, there is a number of sources that do not
match within this scenario, 8 type 1 AGN are absorbed (3\% total) and
6 type 2 AGN are unabsorbed (20\% total). Nonetheless, for half of the
absorbed type 1 AGN, the errors in the intrinsic column density are
consistent with these sources being unabsorbed. It is important to
note that the expected intrinsic column density derived from optical
reddening depends on the assumed gas-to-dust ratio. If the intrinsic
gas-to-dust ratio differs significantly from the Galactic value (the
standard gas-to-dust ratio that is usually used), small differences
between the expected value from optical observations and the measured
value from X-rays are expected. Three of the unabsorbed type 2 AGN are
consistent with being absorbed within errors, but only by low amounts
of intrinsic absorption ($\le$ 10$^{22}$ cm$^{-2}$). This implies that
the optical/X-ray type mismatch could be more common for sources that
are optically classified as type 2 AGN. However, it is difficult to
quantify how frequent the ``mismatches'' between optical and X-ray
classification are.

In an X-ray selected sample, the least biased way to estimate the
exceptions to the unified models is to compute the fraction of
optically classified type 1/2 AGN among the absorbed/unabsorbed
sources. Elusive AGN have to be removed because their classification
is derived from X-rays. We also excluded those absorbed/unabsorbed
sources that could be unabsorbed/absorbed within errors. In this way,
we found that there are only four type 1 AGN among ``truly'' absorbed
sources and two type 2 AGN among ``truly'' unabsorbed
sources. Therefore, the resulting fraction of exceptions to the
unified models for the XBS AGN sample turns out to be of $\sim$
3\%. For the HBSS only, which is almost completely identified, we find
that the fraction of type 2 AGN among unabsorbed sources is only
3\%. A similarly low value of 1\% is obtained for the BSS. For the
fraction of type 1 AGN among absorbed sources, we find 17\% for the
HBSS, whereas for the BSS it turns out to be $\sim$ 30\%. However, if
we take into account the larger number of unidentified sources on the
BSS, seven of which are probably absorbed AGN (see discussion at the
end of this section), this number could decrease to match the one
obtained for the HBSS.

To compare our findings with previously reported results, we selected
a 10$^{22}$ cm$^{-2}$ limit, which is the one that is usually used by
other authors, to separate between absorbed and unabsorbed
sources. Using this limit and considering the total XBS sample, we
computed a fraction of unabsorbed type 2 AGN among the total number of
type 2 AGN of $\sim$ 36\%. This number agrees with reported values of
unabsorbed type 2 AGN shown in \citet{panessa02} and \citet{akylas09}
($\sim$ 10-30\% and $\sim$ 20\%, respectively). However, it has to be
pointed out that none of the reported fractions in \citet{panessa02}
and \citet{akylas09} has been derived from complete samples. For
example, if we consider only the HBSS, which is almost completely
identified and that it is less biased against absorbed sources, the
value decreases from 36\% to 20\%; this is expected to be a more
reliable fraction than that computed using the total XBS sample.
Moreover, errors in the resulting N$_{H}$ values are not usually
considered either. If we remove unabsorbed type 2 AGN that could be
absorbed within 90\% confidence errors, the fraction decreases to
$\sim$ 5\%. Therefore, caution must be exercised when computing the
fraction of exceptions to unified models; these fractions have
probably been overestimated in the past.

The existence of unabsorbed type~2 AGNs has no clear explanation so
far. Some recent models \citep{elitzur06,nicastro00} show that the BLR
could not form under particular condition. For instance, it has been
proposed that the BLR may disappear below bolometric luminosities of
$\sim$ 10$^{42}$ erg s$^{-1}$ (\citet{elitzur06}) or below a critical
accretion rate (L$_{bol}$/L$_{Edd}$ $\sim$ 1-4$\times$10$^{-3}$ for
SMBH masses ranging from 10$^{6}$ to 10$^{9}$ solar masses; where
L$_{bol}$ and L$_{Edd}$ are the bolometric and Eddington luminosities,
respectively, \citealt{nicastro00}).  Nevertheless, the range of
luminosity and accretion rates covered by the unabsorbed type 2 AGN in
our sample (L$_X$ from 10$^{42}$ to $\sim$ 2$\times$10$^{44}$ erg
s$^{-1}$ and accretion rates from $\sim$ 10$^{-3}$ up to $\sim$1,
Caccianiga et al. in preparation) make these interpretations not
applicable to the sources of the XBS sample.

A possible alternative explanation is that unabsorbed Type 2 AGN are
indeed Compton-tick (CT) i.e. sources where the amount of intrinsic
absorption is so high (above 10$^{24}$ cm$^{-2}$) that the absorption
cut-off falls outside the observed spectral range. Using X-ray data
limited in the 2-10 keV energy band, it would not be possible to
compute the actual column density and we would end up with an
optically type~2 AGN with no sign of absorption in the X-rays. The
unabsorbed type~2 AGN in our sample are XBSJ012057.4--110444,
XBSJ031146.1--550702, XBSJ100032.5+553626, XBSJ141235.8--030909,
XBSJ230522.1+122121 and XBSJ221951.6+120123.  XBSJ100032.5+553626 is
an elusive AGN, whose Compton-thick nature was studied and discarded
as a possible explanation in \citet{caccianiga07}. To test the
Compton-thick hypothesis for the remaining five unabsorbed type 2 AGN,
we used the diagnostic diagram by \citet{bassani99}, which make use of
the thickness parameter (T) and the Fe K$\alpha$ line EW to separate
Compton-thick from Compton-thin sources. The thickness parameter
represents the ratio between the 2-10 keV observed flux (corrected for
Galactic absorption) and the reddening-corrected flux of the
[OIII]$\lambda$5007$\AA$ emission line. Compton-thick sources usually
locate at T $<$ 1 and large Fe K$\alpha$ equivalent widths.  The
values for the [OIII] fluxes used here were computed following the
prescription and assumptions discussed in \citet{caccianiga07}.
Regarding the Fe K$\alpha$ EWs, only upper limits (at the 90$\%$
confidence limit) could be derived; we assumed in all cases an
unresolved neutral emission line centered at 6.4 keV.

\begin{figure}[ht]
  \centering
  \includegraphics[angle=-90,width=9cm]{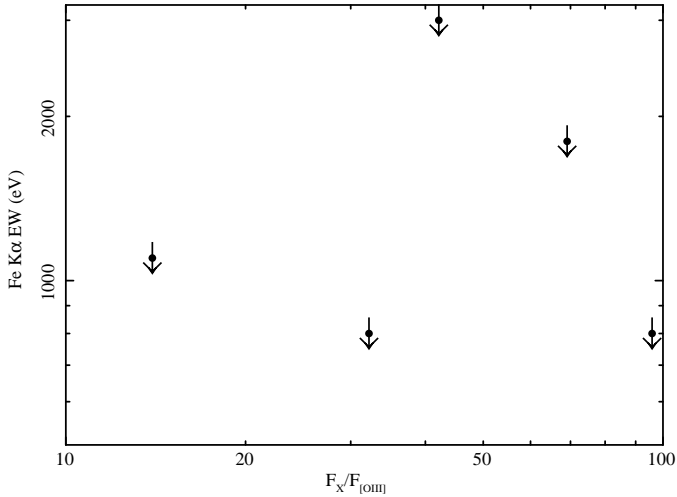}
  \caption{Fe K$\alpha$ EW versus the thickness parameter T = F$_{X}$/F$_{[OIII]}$ for the 5 unabsorbed type 2 AGN. Vertical arrows correspond to upper limits  on the Fe K$\alpha$ EW.}
  \label{bassani_diag}
\end{figure}

Our results for the five sources in consideration here are plotted in
Fig.~\ref{bassani_diag}; obviously all our sources are well above the
F$_{X}$/F$_{[OIII]}$ = 1 limit. The computed [OIII] fluxes are not
corrected for extinction from the host-galaxy. Correcting for the
host-galaxy extinction would increase the [OIII] fluxes thus
decreasing the resulting T values. Nonetheless, the maximum expected
extinction (A$_{V}$ $\sim$ 1 for galaxies with ongoing intense
star-formation, \citet{cal99}) is not high enough to place these
sources below the T = 1 limit.  We conclude that the CT hypothesis is
not a valid explanation for the unabsorbed type~2 AGN in the XBS
sample. Further investigations are thus required to understand the
nature of these intriguing sources.

\begin{figure}[ht]
  \centering
    \includegraphics[angle=0,width=9cm]{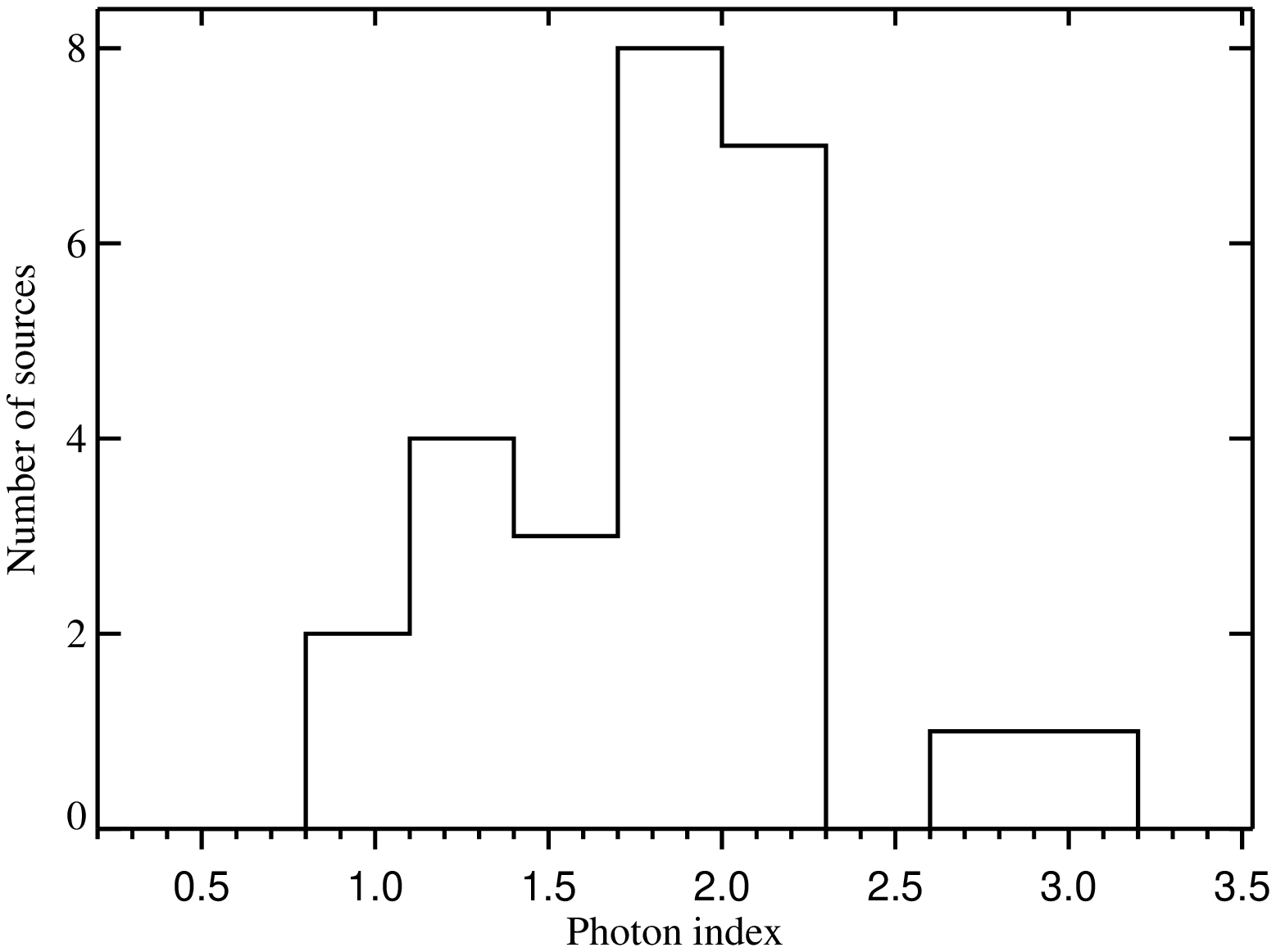}
  \caption{Photon index distribution for the 27 unidentified sources
    in the XBS sample.}
  \label{unid}
\end{figure}

Finally the observed fractions of absorbed AGN (number of absorbed AGN
to total number of AGN) are $\sim$ 35\% for the HBSS and $\sim$ 7\%
for the BSS (34\% and 6\%, respectively, if we do not take into
account those sources that are consistent with being unabsorbed within
errors). This difference is expected because the selection at higher
energies is less affected by obscuration, therefore the HBSS is much
more efficient in selecting absorbed sources. To investigate how the
larger number of unidentified sources in the BSS could affect this, we
performed a simple power law X-ray spectral fit over the unidentified
AGN in the XBS sample. The resulting photon index distribution is
shown in Fig.~\ref{unid}. We can see that seven (including the two
unidentified sources in the HBSS) out of the 27 unidentified sources
display a photon index below 1.5. If these seven sources turn out to
be absorbed AGN, the fraction of absorbed AGN in the BSS sample would
increase to 10\%.

\subsection{Hardness ratios}

\begin{figure*}[ht]
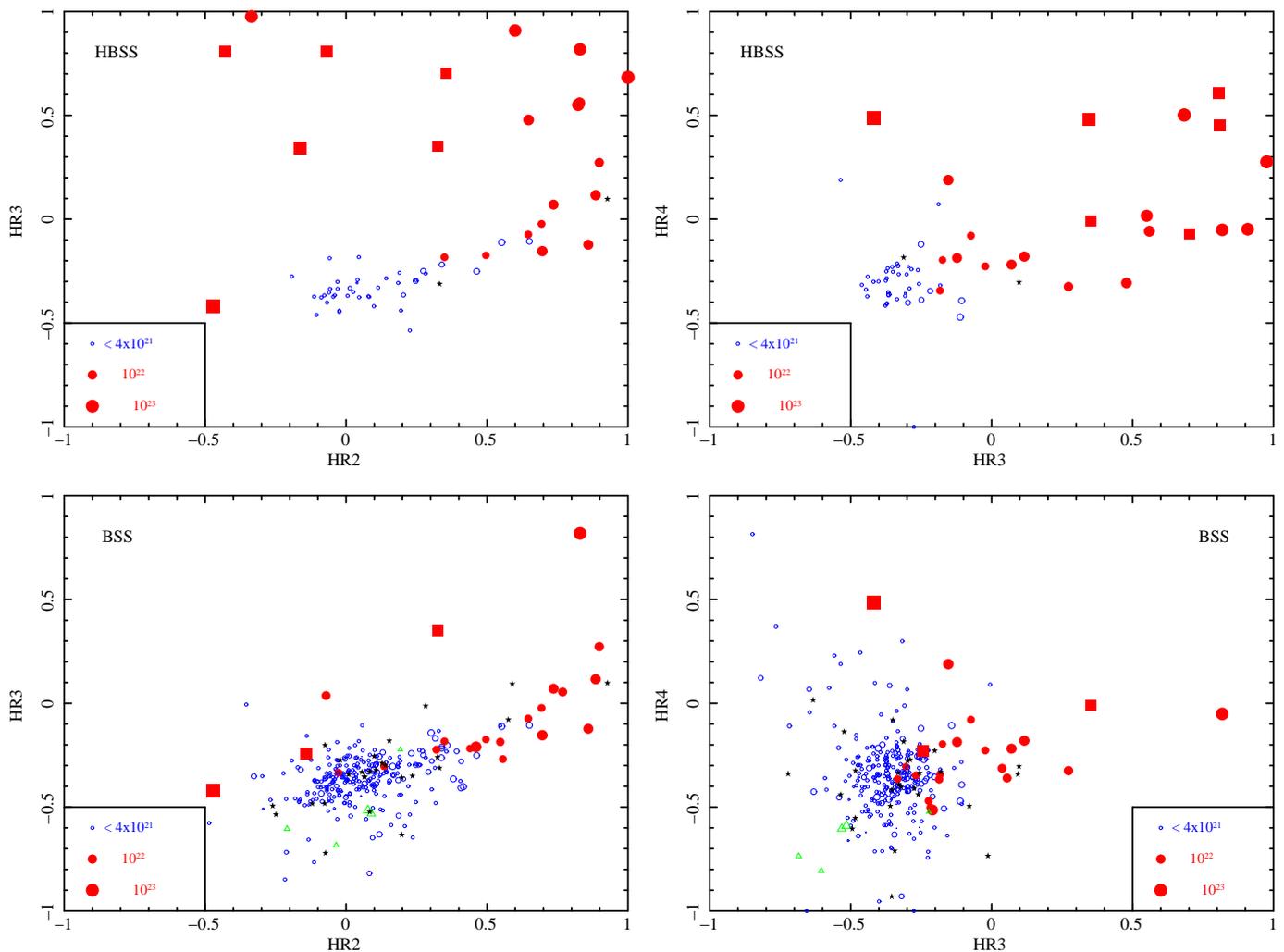

  \centering
$$
  \begin{array}{cc}
    \includegraphics[angle=-90,width=9cm]{15227fg5a.ps} & \includegraphics[angle=-90,width=9cm]{15227fg5b.ps}\\
    \includegraphics[angle=-90,width=9cm]{15227fg5c.ps} & \includegraphics[angle=-90,width=9cm]{15227fg5d.ps}
  \end{array}
$$
  \caption{Hardness ratios for the unabsorbed (open circles) and absorbed (filled circles) identified sources corresponding to the HBSS (top panels) and the BSS (bottom panels), the circle sizes are proportional to the measured intrinsic absorption. Stars correspond to unidentified sources. Filled squares correspond to sources whose best fit is a leaky model.}
  \label{hr}
\end{figure*}

Another way to estimate the source type by using X-rays is to compute
the hardness ratios (X-ray colors). We extracted the hardness ratios
HR2, HR3, and HR4 from the 2XMM-Newton
catalog\footnote{http://xmmssc-www.star.le.ac.uk/Catalogue/2XMM/}
\citep{watson09}, which are defined as

\begin{equation}
HRn=\frac{CR_{n+1}-CR_{n}}{CR_{n+1}+CR_{n}},
\end{equation}

where CR$_{n}$ is the ``vignetting''-corrected count rate in the
energy band n. The 2, 3, 4, and 5 energy bands correspond in our case
to the count rates in the 0.5-1.0, 1.0-2.0, 2.0-4.5, and 4.5-12.0 keV
energy bands, respectively. To compare them with the unidentified
sources, the hardness ratios for the identified AGN are plotted along
with the ones for the unidentified sources in Fig.~\ref{hr}. For
clarity errors are not plotted in this figure given the large number
of sources. To see how the hardness ratios relate to the fitted
absorption, we made the symbol sizes proportional to the measured
intrinsic column density. Filled squares refer to sources for which
the best-fit model is a leaky model. If we do not take into account
these latter sources, we can see how the most absorbed sources
concentrate in the upper right in the HR3 vs. HR2 figures. For the
HBSS, one of the unidentified sources is clearly within that region,
whereas the remaining one is not, although it could be moderately
absorbed given its photon index from the power-law fit. For the BSS,
only about two lie within that region. Evidently also the intrinsic
absorption seems to increase as the hardness ratio HR3 increases.

\begin{figure}[ht]
  \centering
    \includegraphics[angle=-90,width=9cm]{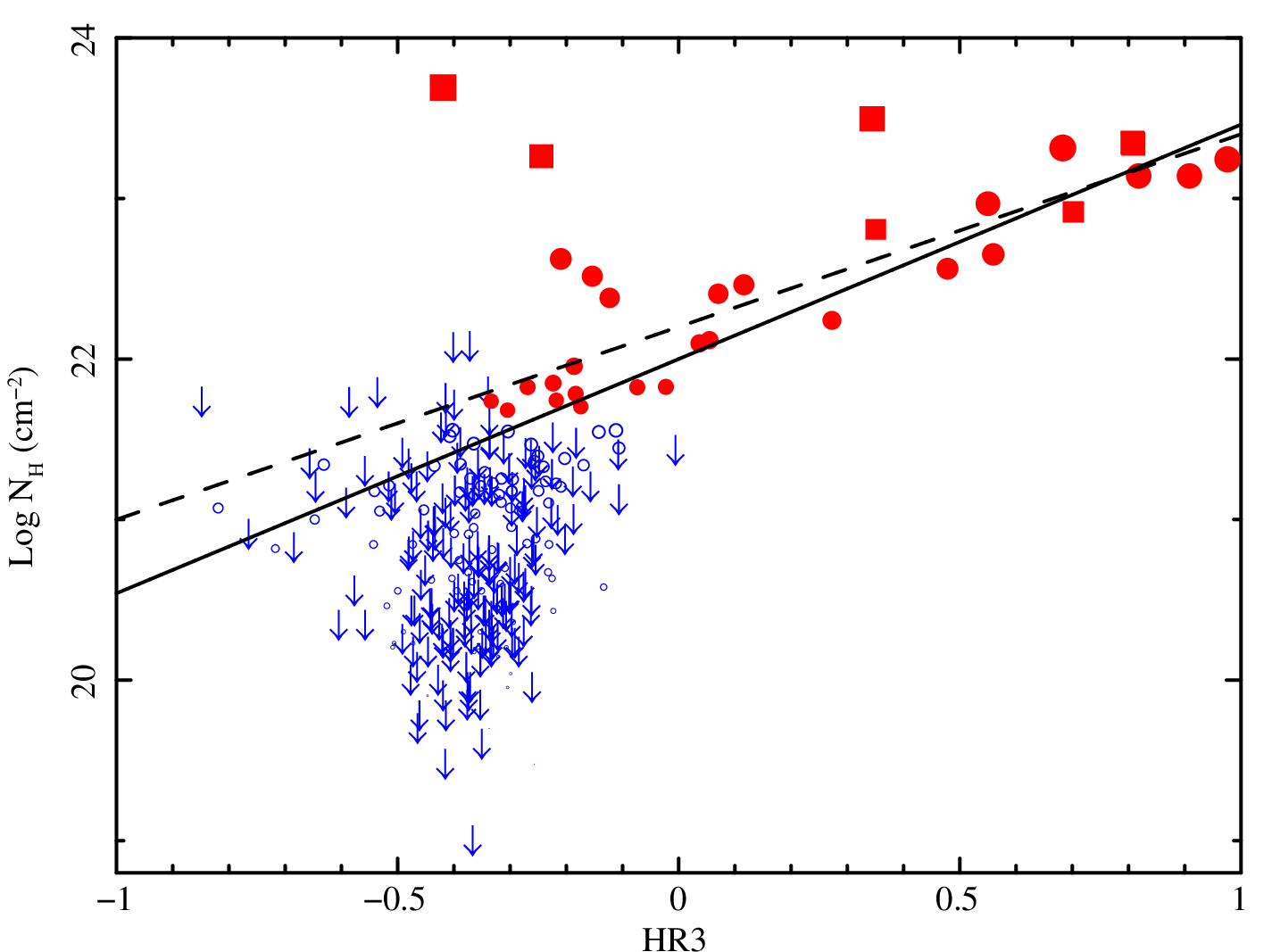}
  \caption{Intrinsic column density against HR3 for unabsorbed (open circles) and absorbed (filled circles) sources. Filled squares correspond to sources whose best fit is a leaky model. Dashed and solid lines correspond to different fits to the observed correlation (see text for details).}
  \label{nhvshr3}
\end{figure}

In Fig.~\ref{nhvshr3} the measured column densities are plotted
against HR3. Obviously there seems to be a correlation between the
amount of intrinsic absorption and HR3 for absorbed sources, again not
taking into account sources with a leaky shape. To derive an estimate
of the intrinsic column density for unidentified sources, we fitted a
linear model to this correlation. In this way, an intrinsic column
density can be estimated even when the X-ray data quality is too poor
to carry out a reliable spectral analysis. We performed the fit in two
different ways. The first one was to fit a linear model by using
$\chi^{2}$ statistics. We considered only those sources with a
detected value of the intrinsic column density higher than
4$\times$10$^{21}$ cm$^{-2}$, i.e. absorbed sources, and we excluded
sources with a leaky shape. This selection criterion resulted in a
total of 25 sources. With a Spearman rank correlation analysis we
confirmed a strong correlation ($\rho$=0.82, probability=0.0001). The
resulting fitted relation corresponds to the dashed line in
Fig.~\ref{nhvshr3} and it is

\begin{equation}
$$
\label{eq1}
{\rm Log(N_{H})}=22.2(\pm0.2)+1.2(\pm0.2){\rm HR3}.
$$
\end{equation}

Our second approach was to use all sources with a column density or
upper limit above the 4$\times$10$^{21}$ cm$^{-2}$ threshold and also
all unabsorbed sources whose intrinsic column densities values were
consistent with this limit within 90\% errors, which were 85 sources
in total. To perform this analysis, we used the ASURV package ({\it
  Astronomy Survival Analysis}, \citealt{iso90}, which implements the
methods presented in \citealt{iso86}). Applying a Spearman rank
analysis, including the upper limits, we found again that there is a
strong correlation between the intrinsic column density and HR3
($\rho$=0.84, probability$<$0.0001). We performed linear regression
with the parametric EM algorithm, solid line in Fig.~\ref{nhvshr3},
which assumes Gaussian residuals as in $\chi^{2}$ statistics,
obtaining the relation

\begin{equation}
$$
\label{eq2}
{\rm Log(N_{H})}=22.00(\pm0.04)+1.46(\pm0.10){\rm HR3}.
$$
\end{equation}

Both relations in Eq.~\ref{eq1} and \ref{eq2} give similar estimates
for the intrinsic column density and can be applied up to redshift
$\sim$ 1, given the energy bands considered in the computation of HR3,
and HR3 $>$ 0. Making use of these relations, an estimate of the
intrinsic column density can be obtained for unidentified sources with
small number of collected counts in X-ray surveys. As an example, we
found for our unidentified sources that about eight out of the 27
unidentified sources could be absorbed AGN, consistent with what we
obtained from the spectral analysis.

\section{Photon index}
\begin{figure*}[ht]
  \centering
$$
  \begin{array}{cc}
    \includegraphics[angle=0,width=9cm]{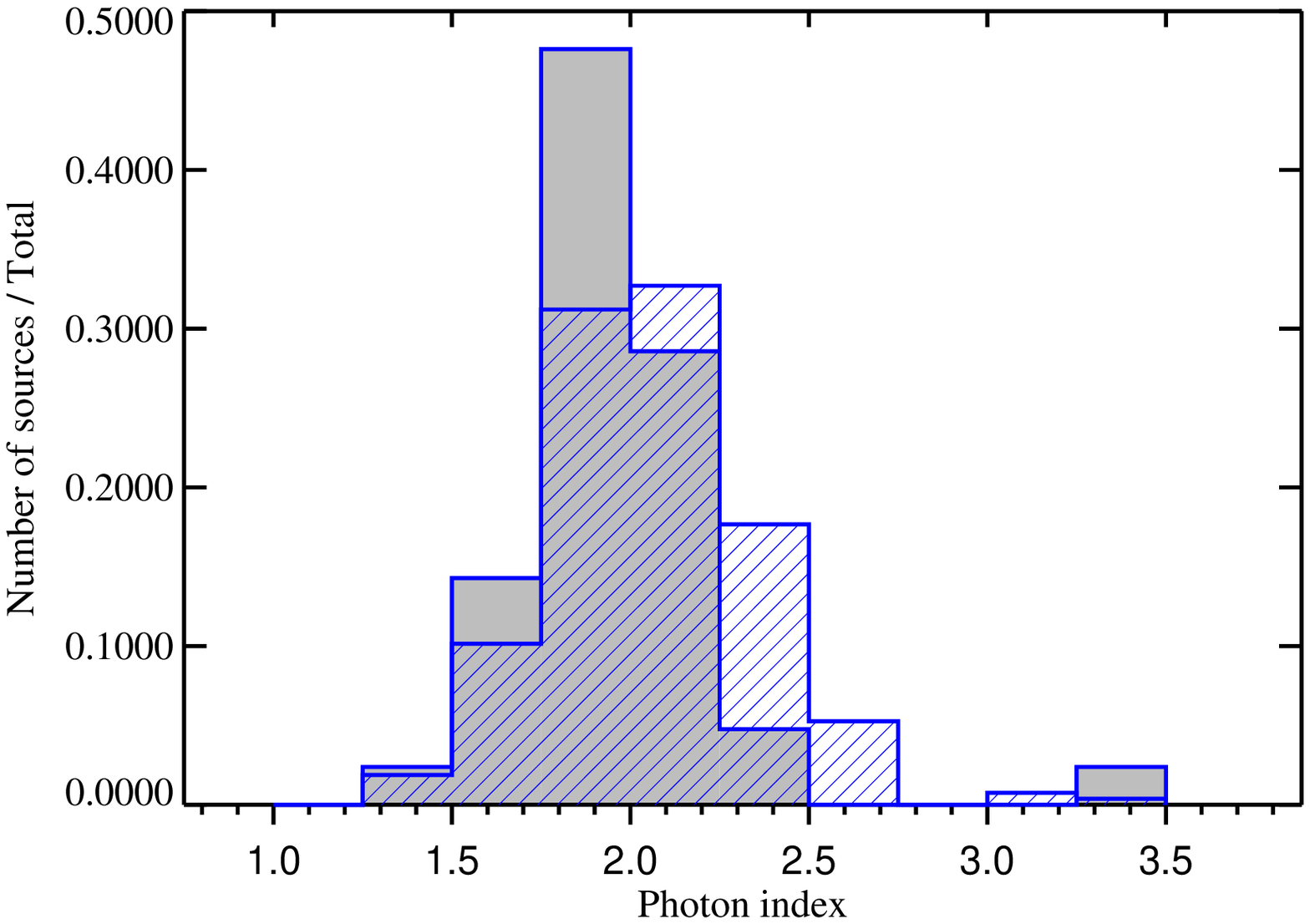} & \includegraphics[angle=0,width=9cm]{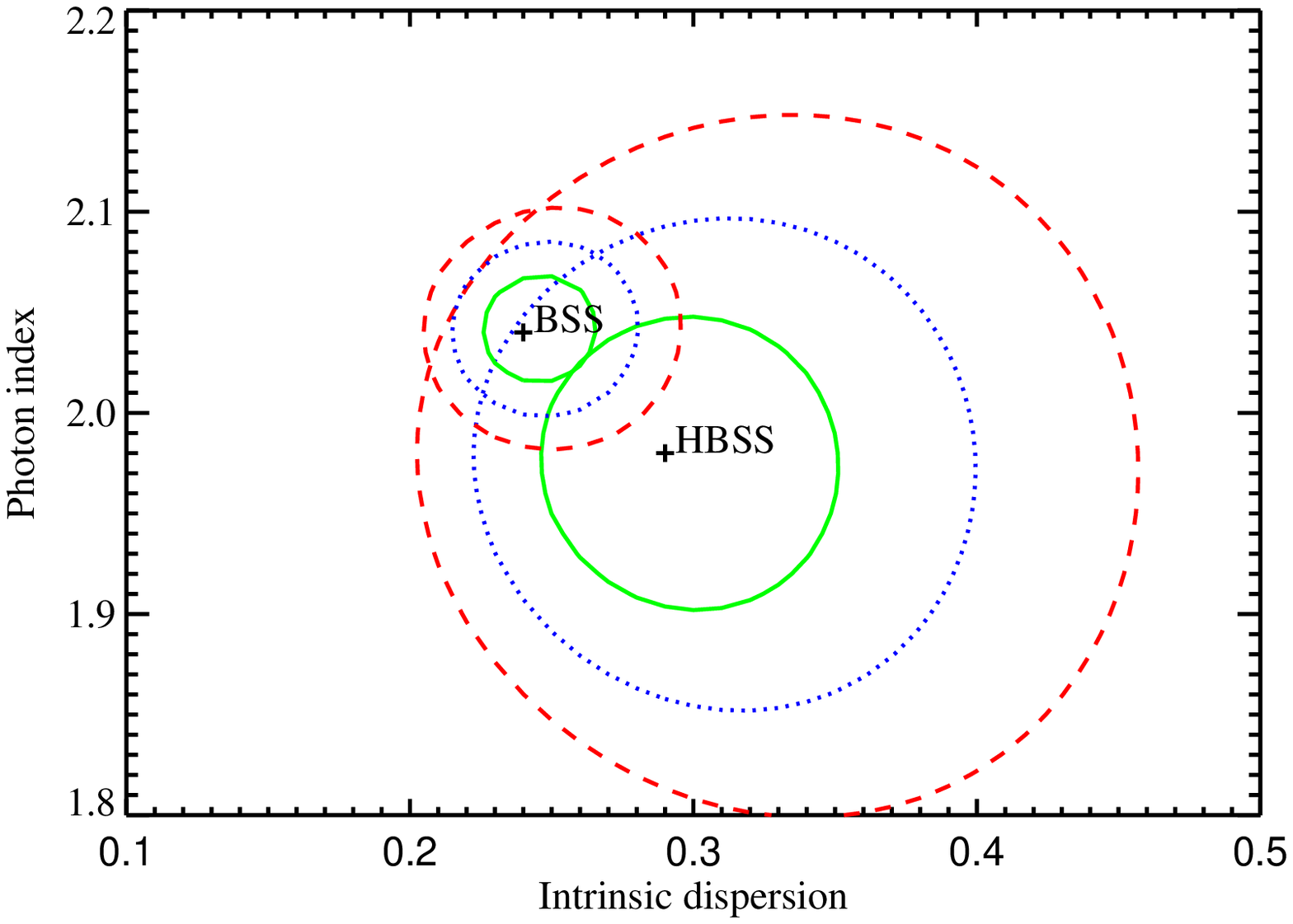}\\
  \end{array}
$$ 
 \caption{Left panel: Photon index distribution for the BSS (dashed
    histogram) and the HBSS (filled histogram). Right panel: Computed mean
    photon index and intrinsic dispersion for the BSS and the HBSS
    along with the 1$\sigma$ (solid line), 2$\sigma$ (dotted line)
    and 3$\sigma$ (dashed line) confidence contours.}
  \label{gammadist}
\end{figure*}

Unfortunately, for half of type 2 AGN (14 out of 29) the photon index
was fixed to 1.9 during the spectral fit to better constrain the
intrinsic absorption, so we restricted our analysis of the power law
index to those type 1 AGN in the XBS AGN sample for which we were able
to measure $\Gamma$, 267 type 1 AGN. We computed its mean and its
intrinsic dispersion making use of \citet{mac88} likelihood
maximization technique. Because the errors on the photon index are not
symmetric, we used the average value for the individual errors in each
case. We obtain a mean value of $\langle\Gamma\rangle$=2.05$\pm$0.03
with an intrinsic dispersion of $\sigma$=0.26$\pm$0.02 for the BSS and
$\langle\Gamma\rangle$=1.98$\pm$0.08 with an intrinsic dispersion of
$\sigma$=0.29$\pm$0.05 for the HBSS, in agreement with previous works
(\citealt{mateos10}, \citealt{young09}, \citealt{dadina08},
\citealt{mai07} \citealt{page06}, \citealt{mateos05a},
\citealt{mateos05b}). Errors were extracted from the 1$\sigma$
confidence contours. The measured power-law index distribution for the
BSS and the HBSS is shown in Fig.~\ref{gammadist} along with the
computed mean and intrinsic dispersion and their confidence
contours. There are three type 1 AGN for which the resulting photon
index is $>$ 3. Two of them are NLSy1s (narrow line Seyfert 1
galaxies), known to show these high values for the photon index, and
the remaining one is a Seyfert galaxy with a low number of counts in
its X-ray spectra.

\begin{figure*}[ht]
  \centering
$$
\begin{array}{c c}
   \includegraphics[angle=0,width=9cm]{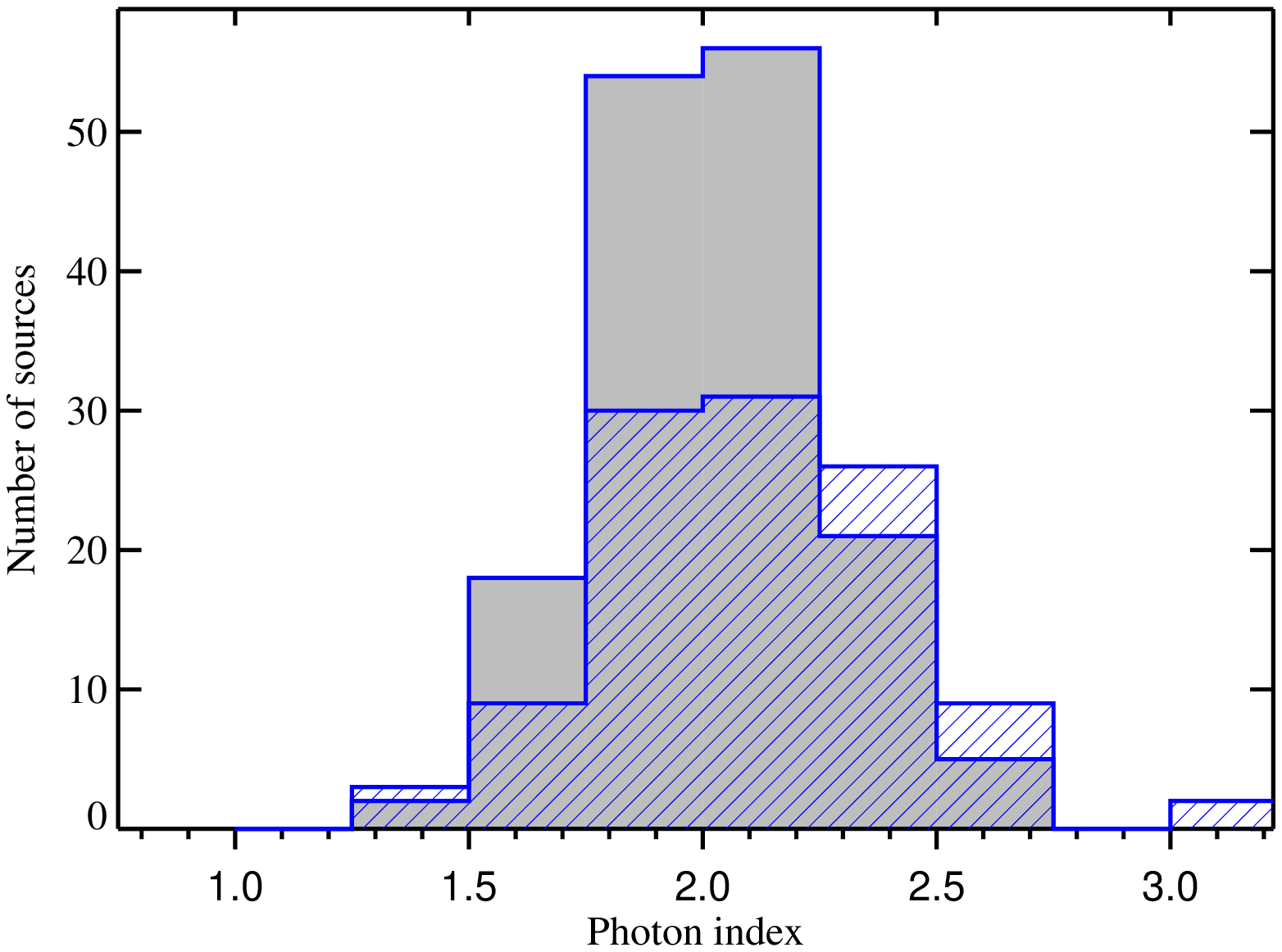} & \includegraphics[angle=0,width=9cm]{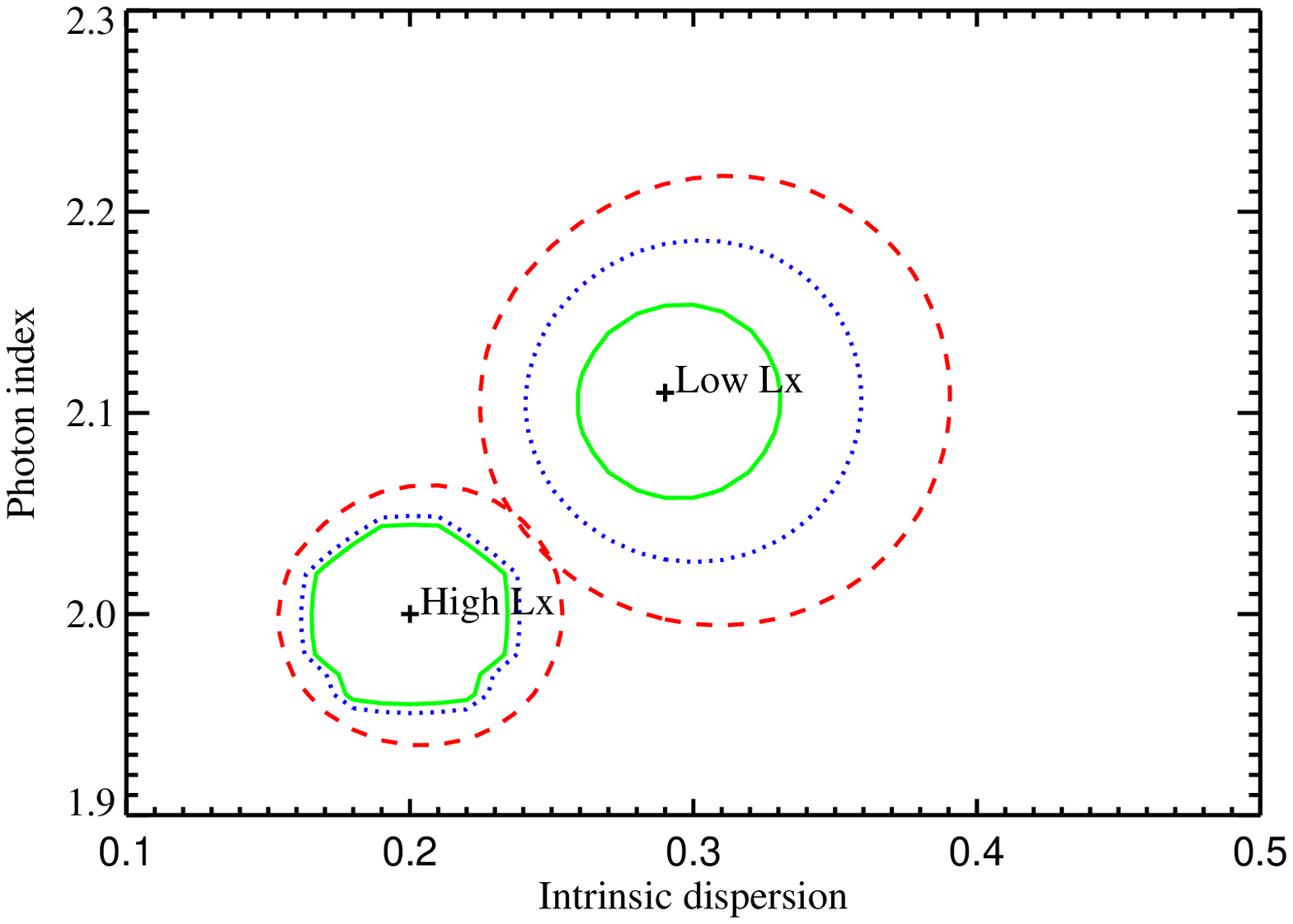}\\
\end{array}
$$
  \caption{Left panel: Photon index distribution corresponding to the
    low-luminosity (dashed histogram) and the high-luminosity (filled
    histogram) subsamples. Right panel: Computed mean and intrinsic
    dispersion for the low- and high-luminosity subsamples along with
    the 1$\sigma$ (solid line), 2$\sigma$ (dotted line) and 3$\sigma$
    (dashed line) confidence contours.}
  \label{lxgamma}
\end{figure*}

\begin{figure*}[ht]
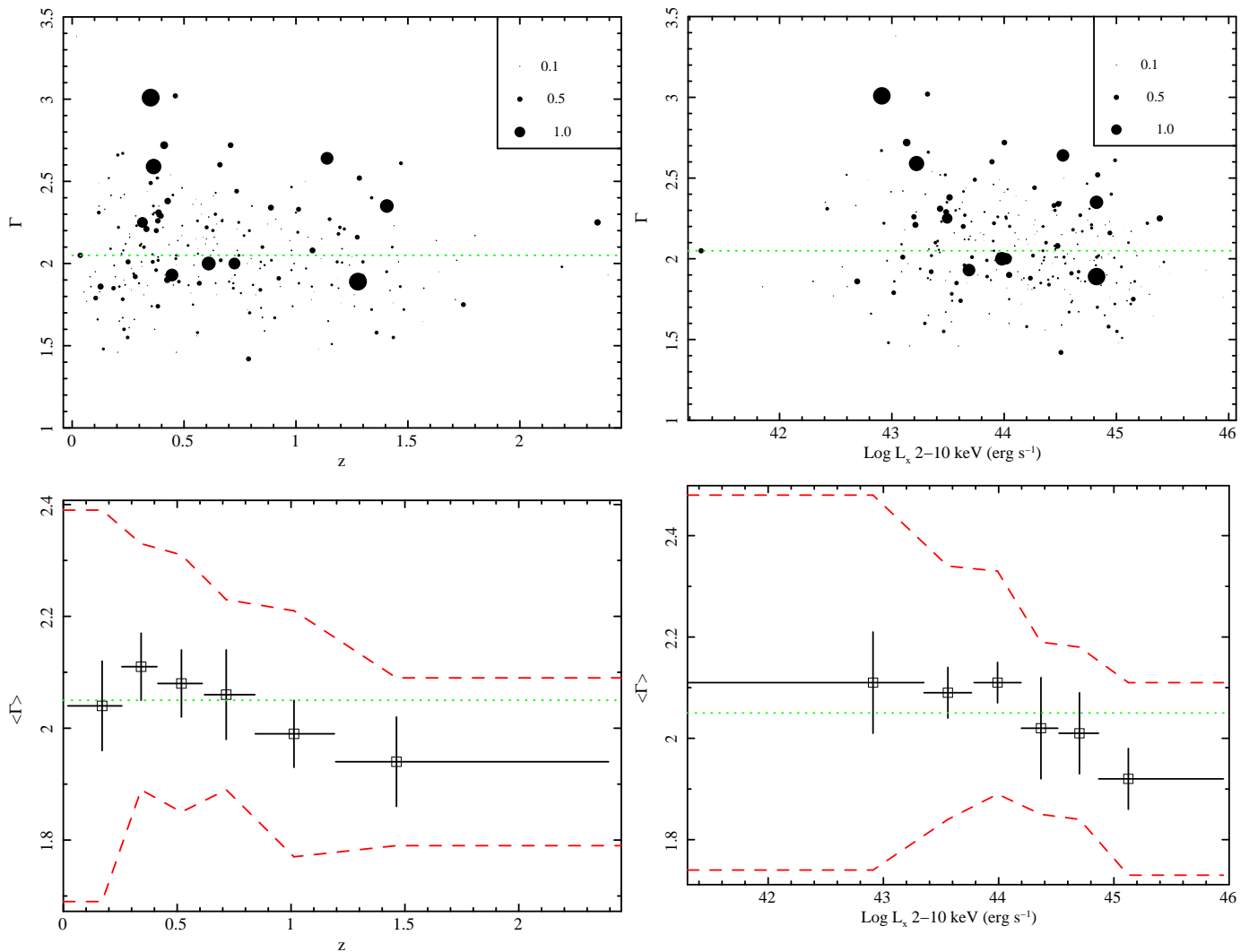

  \centering
$$
  \begin{array}{cc}
    \includegraphics[angle=-90,width=9cm]{15227fg9a.ps} & \includegraphics[angle=-90,width=9cm]{15227fg9b.ps}\\
    \includegraphics[angle=-90,width=9cm]{15227fg9c.ps} & \includegraphics[angle=-90,width=9cm]{15227fg9d.ps}\\
  \end{array}
$$
  \caption{Top panel: Type 1 AGN photon index versus redshift (left)
    and luminosity (right). The size of the circles indicates the size
    of the errors on the photon index. The horizontal dotted line
    corresponds to the average photon index for the whole
    sample. Bottom panel: Type 1 AGN average photon index versus
    redshift (left) and luminosity (right) for the luminosity and
    redshift bins. The error bars correspond to the mean error at
    1$\sigma$ confidence level, whereas the dashed lines mark the
    values for the intrinsic dispersion at each bin.}
  \label{gamma}
\end{figure*}

To check for possible dependence of the photon index on redshift or
luminosity, the values for the individual photon index measurements
are presented along with the source redshift and luminosity in
Fig.~\ref{gamma}. We find an anti correlation between the power-law
index and the X-ray luminosity ($\rho$=-0.21,
probability=8$\times$10$^{-4}$) and marginally redshift ($\rho$=-0.10,
probability=0.09) with the Spearman rank correlation analysis. We also
find a correlation with the 0.5-2 keV flux ($\rho$=0.15,
probability=0.01), which could be caused by undetected intrinsic
absorption as pointed out in \citet{mateos10}. In that work, the
authors found a stronger correlation between the photon index and the
source redshift, but in their case it mainly started above redshift 2
and our sample only contains 3 sources above that redshift. In our
case, we find that the anti-correlation between photon index and
luminosity seems to be the strongest one. Moreover, and given that
this is a flux-limited sample, the dependence on redshift could be
merely caused by the dependence on luminosity. To test this scenario,
we selected two narrow luminosity ranges, 10$^{43}$ to 10$^{44}$ erg
s$^{-1}$, and 10$^{44}$ to 10$^{45}$ erg s$^{-1}$ (below 10$^{43}$ erg
s$^{-1}$ there are not enough sources to perform a reliable analysis),
and applied the same correlation analysis as for the whole type 1 AGN
sample. For the lowest luminosity range, which reaches only z $\sim$
0.8, we find that the anti-correlations turns into a correlation
between the photon index and redshift ($\rho$ = 0.29, probability =
0.006), while for the high-luminosity bin, which reaches z $\sim$ 1.5,
we find that the correlation disappears ($\rho$ = 0.06, probability =
0.69). This may imply that the observed correlation between the source
photon indices and redshifts is mainly driven by an actual correlation
between the photon index and the intrinsic luminosity.

To better explore these correlations, we constructed redshift and
luminosity bins by dividing the sample into six bins with an equal
number of sources (45 sources per bin, 42 in the last bin) and applied
the likelihood-maximization technique to each bin. The results are
presented in Fig.~\ref{gamma}. An anticorrelation between the photon
index and redshift and luminosity seems to be present, but it is
within the intrinsic dispersion at each redshift or luminosity bin. A
similar result is also found in \citet{mateos10}. These authors
pointed out that the hardening of the spectra at higher luminosities
and redshifts can be caused by a decrement in the detection efficiency
for softer sources and an increment for harder sources, given that the
sample is flux-limited.

To further constrain the correlation of the photon index with the
intrinsic luminosity, which seems to be the strongest correlation in
our case, we divided the sample into two subsamples, the criterion of
which was if the intrinsic luminosity was higher or lower than
10$^{44}$ erg s$^{-1}$. The results are displayed in
Fig.~\ref{lxgamma}. We compute a value of
$\langle\Gamma\rangle$=2.11$\pm$0.04 with an intrinsic dispersion of
$\sigma$=0.29$\pm$0.04 for the low-luminosity subsample and
$\langle\Gamma\rangle$=2.00$\pm$0.05 with an intrinsic dispersion of
$\sigma$=0.20$\pm$0.04 for the high-luminosity subsample. The mean
and intrinsic dispersion results are different almost at the 3$\sigma$
confidence level, although the main difference seems to be on the
intrinsic dispersion. Applying a Kolmogorov-Smirnov (K-S) test, we
find that the probability for both distributions to be drawn from the
same parent distribution is only $\sim$ 1\%. The opposite is found in
\citet{bianchi09b}, who analyzed high-quality X-ray spectral data and
found that Seyfert galaxies show a flatter photon index than quasars.

In principle, the observed correlation of $\Gamma$ with the luminosity
could be owing to the presence of an undetected reflection-emission
component that becomes increasingly important with the luminosity. In
Section~7 we show that this correlation is present also when we
analyze the average spectra, where the reflection component is already
accounted for. This result excludes that the flattening of $\Gamma$
with the luminosity is due to the reflection component.

Because the type 1 AGN sample we used in this particular analysis
could be contaminated by radio-loud (RL) sources, which are expected
to have a flatter photon index on average \citep{ree97,ree00}, we
performed a safety test. Making use of the NVSS/XBS cross-correlation
and analysis presented in \citet{galbiati05}, we removed all RL
sources within these 267 type 1 AGN, which were 14 sources, and
applied the same likelihood analysis by dividing into high- and
low-luminosity AGN. We obtained the same result as for the whole
sample, the only difference was a small decrement on the intrinsic
dispersion for the high-luminosity subsample ($\sigma$=0.19$\pm$0.04),
and the two samples were still different almost at the 3$\sigma$
level.

\section{Soft-excess emission}
We say a source shows a soft-excess emission when the extrapolated
2-10 keV power-law fit displays systematic positive residuals at low
energies. We find that 35 AGN out of the 41 sources that were not
well-fitted by an absorbed power law display a soft-excess. For 29 out
of these 35 AGN, we are able to find an additional component that
significantly improves the simple power-law fit, as measured by
F-test. Assuming a fraction of spurious detections of 0.05, given our
F-test significance limit of 95\%, this number corresponds to
5$^{+2}_{-4}$\% of the total XBS AGN sample (14$^{+9}_{-7}$\% for the
HBSS and 4$^{+2}_{-2}$\% for the BSS). If we only take into account
sources below z=0.5 (beyond that value most of the soft-excess
emission is redshifted outside the EPIC energy range) the fraction of
sources increases to 11$^{+6}_{-5}$\% for the XBS sample
(20$^{+12}_{-10}$\% for the HBSS and 9$^{+6}_{-4}$\% for the BSS), a
value closer to the reported values in recent works
\citep{mateos10,bianchi09b}, although still lower. It should be noted
that our computed value has to be considered as a lower limit because
we did not search for soft-excess emission for all our sources, but
only for the ones for which a simple power law gives a probability $<$
10\%. This could also explain why the fraction of sources showing
soft-excess is larger for the HBSS than for the BSS. The difference in
the fraction of detected soft-excess in both samples is likely caused
by differences in the data quality. The collected number of counts for
the HBSS spectra is larger on average than for the BSS. As the
spectral quality increases, it becomes easier to detect and
characterize additional components. Undetected soft-excess would also
increase the value for the measured photon index, and this in turn
could be contributing to increase the computed average photon index
for the BSS and resulting in a higher value than the one for the HBSS.

In the case of absorbed AGN and thanks to Chandra and {\it XMM-Newton}
grating spectra, this soft-excess is known to be associated to
scattered emission hundreds of pc far form the central source, likely
by the NLR clouds (see for example \citealt{bianchi06}). Indeed, all
absorbed AGN that display soft-excess, five type 2 and two type 1 AGN,
are best-fitted either by a leaky model or by a power law plus a
thermal component that could arise from the host galaxy given its low
luminosity.

The case of unabsorbed AGN is more complex. Soft-excess emission has
usually been attributed to the hard tail of the thermal emission from
the accretion disk or to optically-thick comptonization of EUV disk
photons \citep{ross92, shimura93}, but these models are unable to
explain either the higher temperatures usually detected or the fact
that these temperatures seem not to vary with AGN properties such as
the intrinsic luminosity. A recent model, also invoking continuum
emission, explains this soft-excess emission via optically-thin
comptonization of the disk photons \citep{kawa01}, which would explain
the non-dependence on the source luminosity. Two alternative models,
based on atomic processes within the accretion disk, have been
proposed in recent works: the soft-excess emission could come form
relativistically blurred reflection from a partially ionized accretion
disk \citep{crummy06} or from velocity-smeared absorption from
partially ionized material coming from a disk wind \citep{midd07},
although they are indistinguishable at the EPIC energies. Besides, the
quality of our data prevents us from applying them in our spectral
analysis. Using the additional components described in Sect.3, we find
a great variety of best-fit models within the unabsorbed AGN that show
soft-excess in our sample, 27 sources in total: 3 reflection
components, 3 ionized absorbers, 13 black body models and two power
laws plus an absorption edge. For six of them no acceptable fit was
found, therefore the simple power law model was adopted as the
best-fit model.

\begin{figure}[ht]
  \centering
    \includegraphics[angle=-90,width=9cm]{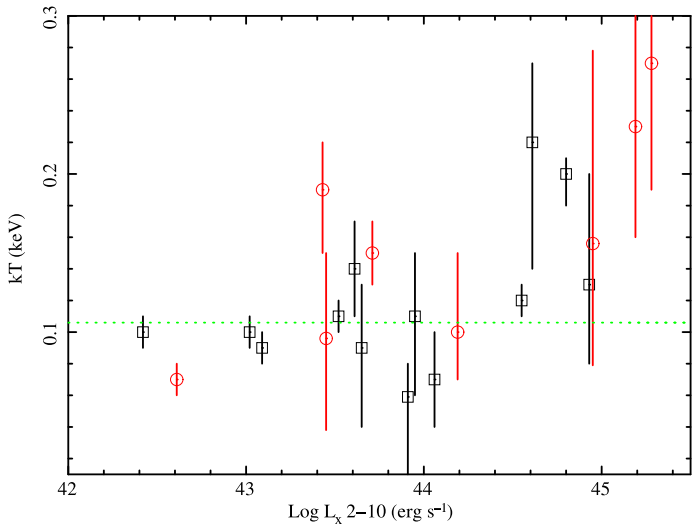}
  \caption{Black body temperature versus 2-10 keV luminosity for
    unabsorbed AGN. Squares correspond to the temperature values if a
    power law plus a black body component is considered the best-fit
    model, whereas circles correspond to the temperature values in the
    remaining unabsorbed AGN with a soft-excess emission. The
    horizontal dotted line corresponds to the weighted mean.}
  \label{ktvslx}
\end{figure}

As mentioned before, when the soft-excess in unabsorbed AGN is modeled
with a black body model, a value of kT$\sim$0.1 keV is obtained that
does not depend on the source flux, redshift or luminosity. To compare
our results with previous works, we also attempted to fit a black body
plus a power law to all unabsorbed sources with a soft-excess (see
Table~\ref{bbfit}). This improved the simple power-law fit, F-test $>$
95\%, in all but 6 cases, 21 sources in total. The values obtained for
the black body temperature are presented in Fig.~\ref{ktvslx} against
the 2-10 keV luminosity. By using a Spearman rank correlation
analysis, we found a significant correlation between the black body
temperature and the source's luminosity ($\rho$ = 0.60,
probability=0.004). This can be because higher luminosity sources are
at higher redshifts, given that the sample is flux-limited, which
means that the soft-emission is shifted outside the observed energy
range. Therefore, at higher luminosities only black body components of
higher temperatures can be detected. Indeed, if we remove sources at z
$>$ 0.5, the correlation disappears ($\rho$ =0.06, probability =
0.78). Therefore, and given the present statistics, we cannot confirm
if there is an actual correlation between the black body temperature
and the hard X-ray luminosity, although our results suggest that high
black body temperatures can only be reached by sources with high
intrinsic luminosities.

\section{Average spectrum}
A narrow Fe K$\alpha$ emission line (6.4-6.9 keV, depending on
ionization state of the material) is almost ubiquitously observed in
good quality X-ray spectra of AGN \citep{nandra07,bianchi09a}. To
measure a broad and/or a relativistically broadened component requires
an even better quality \citep{gua06}. We searched for this emission
line in our sample, but its presence is only suggested in $\sim$ 20
sources and its parameters are only well constrained in a couple of
cases, all corresponding to a narrow Fe K$\alpha$ line. To improve the
SNR and to detect spectral features that would remain hidden
otherwise, such as the Fe K$\alpha$ line, we averaged the whole sample
following the process described in \citet{corral08}. The basic steps
of this averaging method are

\begin{itemize}
\item Selecting those individual spectra, pn or MOS, with more than 80
  counts in the total 0.2-12 keV EPIC band. For the XBS AGN sample,
  this means to exclude 12 sources, 8 type 1 and 4 type 2 AGN.
\item Fitting pn and MOS spectra for each source individually by using a
  simple absorbed power-law model and leaving the power-law index,
  intrinsic column density, and the normalization as free parameters.
\item Obtaining the incident spectra, i. e. before entering the
  detectors, in flux units (keV cm$^{-2}$ s$^{-1}$ keV$^{-1}$) by
  using the parameters from the previous spectral fit.
\item Correcting for the absorption from our Galaxy and shifting to rest
  frame.
\item Rescaling the individual spectra so that every spectrum has the
  same 2-5 keV rest frame flux.
\item Binning every spectrum to a common energy grid so that the final
  averaged spectrum has at least 1000 counts per bin.
\item Averaging by using a standard mean.
\end{itemize} 

As a final step and to quantify the significance of any spectral
feature, we used simulations: we simulated each source 100 times by
using the fitted model and keeping the same spectral quality as for
the real data. By averaging all the simulations we obtained a
simulated ``continuum'' that should account for the average of
absorbed power laws. Taking one simulation for each real spectrum and
averaging these, we constructed 100 simulated continua from which we
can compute 1$\sigma$ and 2$\sigma$ limits by removing the 32 and 5
extreme values at each bin. In this way, we can say that any excursion
over or below these limits is detected at 1$\sigma$ or 2$\sigma$
confidence level. By using the simulated ``continuum'' and the
confidence limit and comparing them to our data, we can determine if
there are any significant deviations from a power-law shape and
estimate their significance. Only the energies in the 2 to 15 keV
rest-frame energy band are used in this spectral analysis. For
energies below 2 keV, the averaging method is highly dependent on the
model used to unfold the spectra, and for energies above 15 keV noise
becomes too important. The resulting best-fit models and parameters
for each case are shown in Table \ref{averagefit}.

\begin{figure*}[ht]
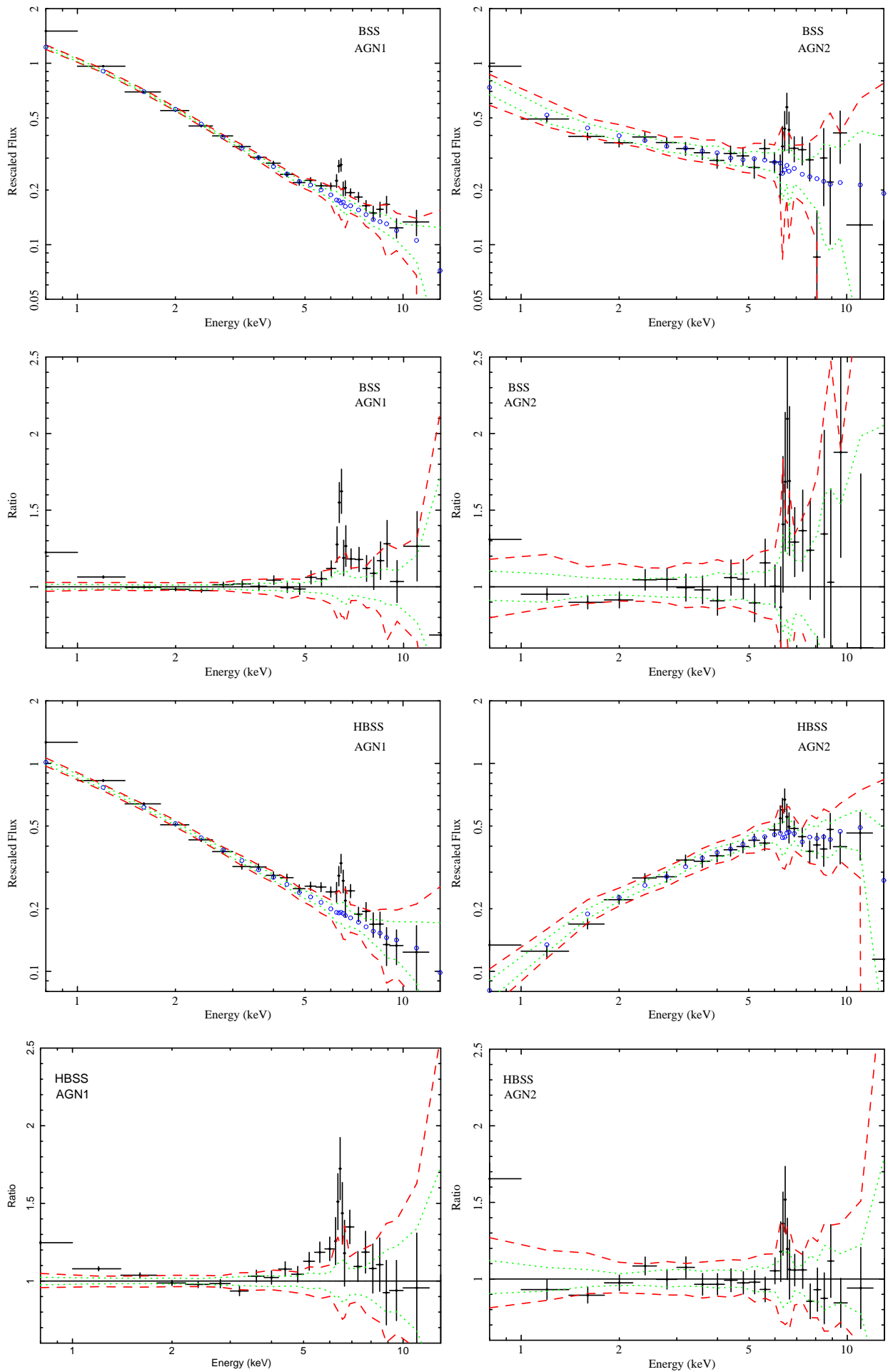

  \centering
$$
  \begin{array}{cc}
    \includegraphics[angle=-90,width=8cm]{15227fg11a.ps} & \includegraphics[angle=-90,width=8cm]{15227fg11b.ps}\\
    \includegraphics[angle=-90,width=8cm]{15227fg11c.ps} & \includegraphics[angle=-90,width=8cm]{15227fg11d.ps}\\
    \includegraphics[angle=-90,width=8cm]{15227fg11e.ps} & \includegraphics[angle=-90,width=8cm]{15227fg11f.ps}\\
    \includegraphics[angle=-90,width=8cm]{15227fg11g.ps} & \includegraphics[angle=-90,width=8cm]{15227fg11h.ps}\\
  \end{array}
$$
  \caption{Type 1 (left) and type 2 (right) AGN average spectrum and average spectrum to simulated continuum ratio (bottom) corresponding to the HBSS and BSS samples. Error bars: real average spectrum, circles: average continuum, dashed line: 2$\sigma$ limit, dotted line: 1$\sigma$ limit.}
  \label{averagehbs}
\end{figure*}

First, and to compare the two samples under study here, we constructed
the average spectra and confidence limits for the type 1 and type 2
AGN within the BSS and HBSS separately. The resulting averaged
spectrum, simulated continuum, and confidence limits for the BSS and
HBSS samples are shown in Fig.~\ref{averagehbs}.

{\bf BSS:} For the type 1 AGN we find that the best-fit model consists
of a power law of $\Gamma$=2.02$^{+0.04}_{-0.03}$ plus a narrow Fe
K$\alpha$ line centered at E=6.40$^{+0.04}_{-0.06}$ keV and equivalent
width EW=110$^{+30}_{-40}$ eV and a reflection component with a
reflection factor R$\sim$0.6. The inclination angle of the reflection
component is always fixed to its default value, i $\sim$ 60 deg,
because it cannot be determined at the same time as the reflection
factor. For the type 2 AGN the best-fit model turns out to be an
absorbed power law ($\Gamma$ fixed to 1.9) with an intrinsic column
density N$_{H}$$<$1.1$\times$10$^{22}$cm$^{-2}$ plus a narrow emission
line (E=6.53$^{+0.14}_{-0.04}$ keV, EW=200$^{+150}_{-150}$ eV) and a
reflection component (R$\sim$1). The fitted intrinsic column density
does not represent the actual average absorption of the sources, but
it is related to the fraction of absorbed sources among the type 2 AGN
as well as their column densities, and as such does not have a
meaningful physical interpretation. In any case, and given that not
only the statistics are much lower for type 2 AGN but also that we are
fitting above 2 keV, the column density cannot be very well
constrained.

{\bf HBSS:} For the type 1 AGN we find that the best-fit model
consists of a power law of $\Gamma$=2.00$^{+0.05}_{-0.07}$ plus a
narrow Fe K$\alpha$ line centered at E=6.44$^{+0.05}_{-0.04}$ keV and
equivalent width EW=80$^{+60}_{-40}$ eV and a reflection component
with a reflection factor R$\sim$0.9. For the type 2 AGN the best-fit
model turns out to be an absorbed power law ($\Gamma$ fixed to 1.9)
with an intrinsic column density N$_{H}$
$\sim$2$\times$10$^{22}$cm$^{-2}$ plus a narrow emission line
(E=6.42$^{+0.09}_{-0.13}$ keV, EW=90$^{+50}_{-60}$ eV) and a
reflection component (R$\sim$1).

The small differences between type 2 AGN average spectra for the BSS
and HBSS are caused by the larger number of absorbed sources for the
HBSS. This is expected because selecting at harder energies makes the
sample less biased against absorbed sources. The relatively low value
for the average column density for the type 2 AGN in the BSS is
consistent with it being due to contribution of unabsorbed type 2 AGN,
which are more numerous in the BSS (which contains six unabsorbed type 2
AGN out of 19 type 2 AGN) than in the HBSS (which only contains one
unabsorbed type 2 AGN out of 20). For both samples and AGN classes,
the detected Fe K$\alpha$ line turns out to be narrow and likely
comes from neutral material, i.e., far from the central source.

\begin{figure*}[ht]
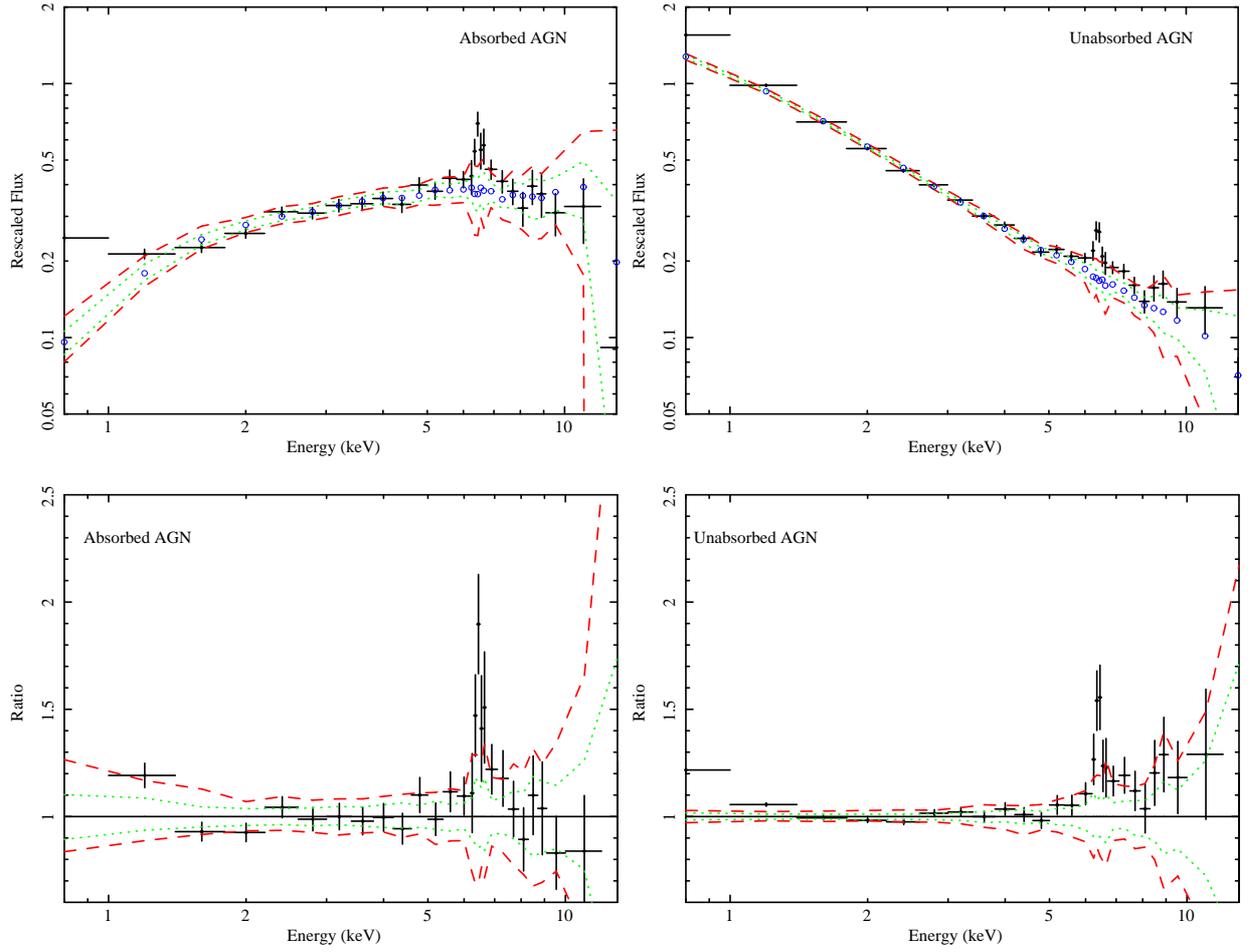

  \centering
$$
  \begin{array}{cc}
  \includegraphics[angle=-90,width=8cm]{15227fg12a.ps} & \includegraphics[angle=-90,width=8cm]{15227fg12b.ps}\\
  \includegraphics[angle=-90,width=8cm]{15227fg12c.ps} & \includegraphics[angle=-90,width=8cm]{15227fg12d.ps}\\
\end{array}
$$
  \caption{Average spectrum (top) and average spectrum to simulated
    continuum ratio (bottom) corresponding to the absorbed (left) and
    unabsorbed (right) sources. Error bars: real average spectrum,
    circles: average continuum, dashed line: 2$\sigma$ limit, dotted
    line: 1$\sigma$ limit.}
  \label{averagexbs}
\end{figure*}

To better characterize the differences in the spectral shape
from absorbed to unabsorbed sources, we divided the whole sample into
absorbed (31 AGN) and unabsorbed (274 AGN) sources and constructed the
average spectra. The results are shown in Fig.~\ref{averagexbs}.

For the {\bf unabsorbed sources} we find that the best-fit model
consists on a power law of $\Gamma$=2.10$^{+0.03}_{-0.03}$ plus a
narrow Fe K$\alpha$ line centered at E=6.40$^{+0.04}_{0.06}$ keV and
equivalent width EW=100$^{+30}_{-40}$ eV and a reflection component
with a reflection factor R$\sim$0.5. A relativistic Fe K$\alpha$ line
is not clearly present and when we attempted to fit one, we did not
obtain a significant improvement, and the line parameters resulted in
unphysical values (such as a very large inclination angle, i $\sim$
80deg, or a extremely high line energy $\sim$ 8 keV). From the
best-fit model, we estimate an upper limit for the EW of a
relativistic Fe K$\alpha$ line contribution of 230 eV at the 3$\sigma$
confidence level for a relativistic emission line \citep{laor}
centered at 6.4 keV and inclination angle of 30deg. For the {\bf
  absorbed sources} the best-fit model turns out to be an absorbed
power law ($\Gamma$ fixed to 1.9) with an intrinsic column density
N$_{H}$$\sim$1$\times$10$^{22}$cm$^{-2}$ plus a narrow emission line
(E=6.47$^{+0.06}_{-0.07}$ keV, EW=100$^{+70}_{-50}$ eV) and a
reflection component (R$\sim$1.5). The main difference between
absorbed and unabsorbed sources seems to be a larger amount of
reflection in the case of absorbed AGN besides the amount of
absorption. This difference is not due to a hidden dependence of the
reflection with luminosity because absorbed and unabsorbed AGN in the
XBS sample display a very similar luminosity distribution with an
average luminosity of $\langle$$L_{X}$(2-10
keV)$\rangle$$\sim$4$\times$10$^{44}$ in both cases and about the same
dispersion. The averaging process used here is designed to study the 2
to 10 keV rest-frame range to study the Fe K$\alpha$ line properties
that minimize the contribution of the underlying continuum and
observational effects. However, for highly absorbed sources (N$_{H}$
$>$ 10$^{23}$ cm$^{-2}$), the way the rescaling is carried out can
give larger weights during the averaging process to the more absorbed
sources. In the final average spectra, this produces a feature that
could mimic the shape of a reflection component (Corral et al. 2011,
in preparation). However, the number of highly absorbed sources is too
small to be responsible for the whole amount of observed
reflection. As a safety test, we removed the eight most absorbed AGN
(N$_{H}$ $>$ 10$^{23}$ cm$^{-2}$) from the average of absorbed
AGN. Given that the remaining number of sources is small in this case,
R cannot be well constrained, but we obtain a lower limit of R $>$ 1.1
at the 90\% confidence level. In summary, although the values obtained
for the reflection components reported here have to be taken as
tentative, the difference in the amount of reflection between absorbed
and unabsorbed AGN seems to be real. For unabsorbed sources, our
results excellently agree with those from studies of local AGN
\citep{nandra07}, from the average of large samples of distant AGN
\citep{chau10} and with the predictions of theoretical models
\citep{bal10}.

\begin{table*}
\caption{Average spectra fit results.}
\label{averagefit}      
\centering
\begin{tabular}{l c c c c c c c}   
\hline\hline             
Sample & N$_{H}$ & $\Gamma$ & R & E & $\sigma$ & EW & $\chi^{2}$/d.o.f \\ 
 & 10$^{22}$ &  & &  &  &  &  \\ 
 & (cm$^{-2}$) &  & & (keV) & (eV) & (eV) &  \\ 
(1) & (2) & (3) & (4) & (5) & (6) & (7) & (8)\\
\hline
\hline
BSS type 1 AGN &  \dots & 2.02$^{+0.04}_{-0.03}$ & 0.6$^{+0.2}_{-0.2}$ & 6.40$^{+0.04}_{-0.06}$ & $<$ 160 & 110$^{+30}_{-40}$ & 21/16\\
HBSS type 1 AGN & \dots & 2.00$^{+0.05}_{-0.07}$ & 0.9$^{+0.2}_{-0.2}$ & 6.44$^{+0.05}_{-0.04}$ & $<$ 140 & 80$^{+60}_{-40}$ & 15/16\\
\hline
BSS type 2 AGN & $<$ 1.1 & 1.9$^{f}$ & 1.0$^{+0.7}_{-0.8}$ & 6.53$^{+0.14}_{-0.06}$ & $<$ 200 & 200$^{+150}_{-150}$ & 13/16\\
HBSS type 2 AGN & 2.0$^{+0.5}_{-0.4}$ & 1.9$^{f}$ & 1.0$^{+0.3}_{-0.3}$ & 6.42$^{+0.09}_{-0.13}$ & $<$ 250 & 90$^{+50}_{-60}$ & 6/16\\
\hline
Unabsorbed AGN & \dots & 2.10$^{+0.03}_{-0.03}$ & 0.5$^{+0.1}_{-0.1}$ & 6.40$^{+0.04}_{-0.06}$ & $<$ 160 & 100$^{+30}_{-40}$ & 18/16\\
Absorbed AGN & 1.2$^{+0.4}_{-0.3}$ & 1.9$^{f}$ & 1.5$^{+0.2}_{-0.3}$ & 6.47$^{+0.06}_{-0.07}$ & $<$ 160 & 100$^{+70}_{-50}$ & 9/16\\
\hline
Low-luminosity type 1 AGN & \dots & 2.11$^{+0.10}_{-0.20}$ & 0.8$^{+0.8}_{-0.5}$ & 6.43$^{+0.06}_{-0.13}$ & $<$ 180 & 110$^{+30}_{-30}$ & 12/16\\
High-luminosity type 1 AGN & \dots & 2.00$^{+0.03}_{-0.03}$ & 0.3$^{+0.1}_{-0.1}$ & 6.39$^{+0.04}_{-0.05}$ & $<$ 130 & 80$^{+30}_{-30}$ & 25/16\\
\hline
\hline
\end{tabular}
\begin{list}{}{}
\item Columns: (1) Sample used to construct the average spectrum; (2) Intrinsic column density; (3) Photon index; (4) Reflection scaling factor; (5) Fe K$\alpha$ central energy; (6) Fe K$\alpha$ width upper limit; (7) Fe K$\alpha$ equivalent width; (8) $\chi^{2}$ to number of degrees of freedom.
\item $^{f}$: Fixed parameter. 
\end{list}
\end{table*}

\subsection{Dependence on redshift and luminosity}

\begin{figure*}[ht]
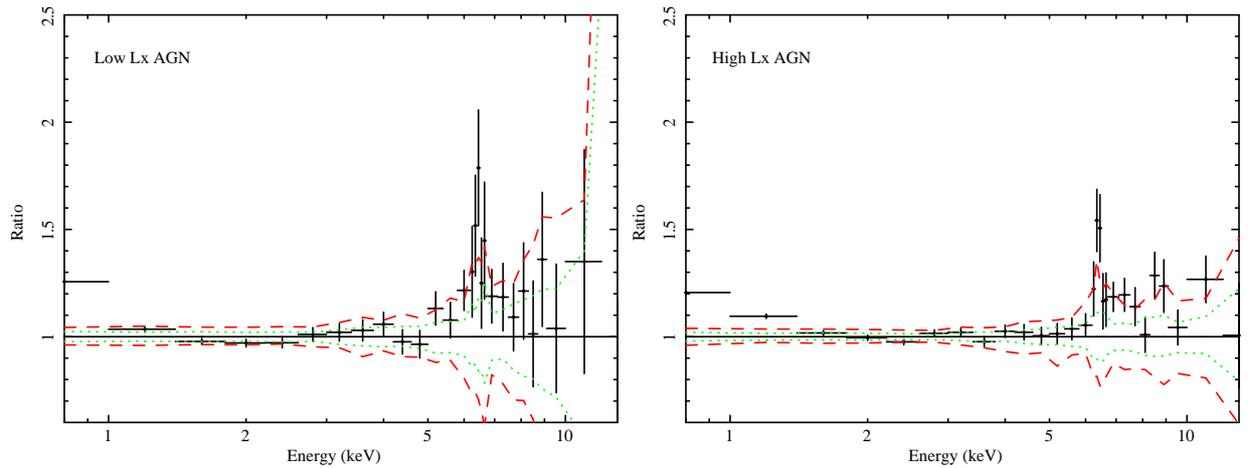

  \centering
$$
\begin{array}{cc}
    \includegraphics[angle=-90,width=8cm]{15227fg13a.ps} & \includegraphics[angle=-90,width=8cm]{15227fg13b.ps}\\
\end{array}
$$ 
 \caption{Average spectrum to simulated continuum ratio corresponding
   to the luminosity subsamples. Error bars: real average spectrum,
   dashed lines: 2$\sigma$ limit, dotted lines: 1$\sigma$ limit.}
  \label{averagelxbins}
\end{figure*}

As we showed in Sect.5, there seems to be a difference between the
spectral shape for low- ($L_x < 10^{44}$ erg s$^{-1}$) and high- ($L_x
> 10^{44}$ erg s$^{-1}$) luminosity type 1 AGN. To explore this
possible difference, we constructed the average spectrum for both
luminosity subsamples. The resulting ratios of the average spectra to
the simulated continua are shown in Fig.~\ref{averagelxbins}. We find
that the best-fit model for both samples consists in a power law with
a narrow Fe K$\alpha$ line centered on $\sim$ 6.4 keV and a reflection
component. Consistently with the results reported in Sect.5 we find
the photon index of the average spectrum for the low-luminosity
subsample to be (marginally) larger than the one for the
high-luminosity subsample. The large error on $\Gamma$ in the
low-luminosity subsample is likely caused by the larger dispersion of
the photon index distribution (see Fig.~\ref{lxgamma}). For the
low-luminosity subsample, the line EW seems to be higher,
EW=110$\pm$30 eV, than for the high-luminosity subsample, EW=80$\pm$30
eV (the so-called Iwasawa-Taniguchi effect, \citealt{iwa}), but both
values are consistent within errors. The resulting reflection fraction
also turns out to be marginally larger for the low-luminosity
subsample, R=0.8$^{+0.8}_{-0.5}$, than for the high-luminosity sample,
R=0.3$\pm$0.1, although it is not well constrained for the former
sample. These results agree with models that predicts a decrease of
the torus covering fraction as the luminosity increases
\citep{lawrence91}, thus decreasing the reprocessing of the radiation
within the torus and also explaining the Iwasawa-Taniguchi
effect. Evidence of this decrement of the covering fraction as a
function of the luminosity have been reported in recent works
(\citealt{maiolino07}, \citealt{dellaceca08}), which point out the
need for the simplest unified schemes to be revised.

We also explored the possible dependence of the spectral shape on
redshift. To this end, we again constructed average spectra by
dividing the sample in different redshift bins. In this case we did
not detect any significant trend of the resulting averaged spectral
shape with redshift. Nevertheless, we point out again that our sample
only reaches redshift $\sim$ 2, and it is above this value where, for
example, \citet{mateos10} found this dependence to become stronger.

\section{Discussion}

In the previous sections we discussed the possible existence of
several statistical correlations. First, we found an anti-correlation
between the photon index and the X-ray luminosity. This correlation is
significant in the analysis of the single spectra, but it is also
marginally present in the analysis of the average spectra. The lower
significance in the latter case is likely caused by the high intrinsic
dispersion in the photon index distribution.  The second correlation,
found in the analysis of the average spectra, is the dependence of the
intensity of the reflection component with both the AGN ``type'' and
luminosity. In particular, the reflection component seems to be
stronger in absorbed AGN and in low-luminosity AGN.

An anti-correlation between the photon index and the X-ray luminosity
has been recently reported by \citet{green09} and \citet{young09},
whereas \citet{saez08} found a ``positive" correlation. It has to be
noted that \citet{saez08} took into account type 1 and type 2 AGN at
the same time, the latter being more numerous than the former, whereas
we here only considered type 1 AGN. The physical explanation for this
anti-correlation is still a matter of debate in the recent
literature. Several authors \citep{shemmer08,risaliti09,grupe10} have
reported a correlation between the photon index and the Eddington
ratio. This dependence could explain the correlations we find if the
Eddington ratio was the actual driver of the luminosity/photon index
anti-correlation. In that case, the low-luminosity subsample could be
sampling a different AGN population or a mixture of very different
accretion states, which could explain the higher dispersion found for
the photon indices. We are currently studying this hypothesis in
deeper detail.

The observed anti-correlation between the reflection component
intensity and the luminosity found in the analysis of the average
spectra confirms similar trends that were already observed in other
samples (e.g. \citealt{nandra97}).  If the observed reflection
component is related to the molecular torus, the trend can be well
explained in the context of the receding torus model according to
which the molecular torus-covering fraction (and thus the intensity of
the reflected component) decreases with the luminosity.
  
On the contrary, the difference seen in the observed reflection
component between absorbed and unabsorbed AGN is, at the moment,
troublesome because it is not easily reconciled with the unified model
(which predicts a larger reflection component for unobscured sources,
see e.g. \citealt{krolik94,murphy09}). It is worth noting that similar
results as those reported here (i.e. more reflection in absorbed
objects) where obtained by \citet{malizia03} and by \citet{deluit03}
from the analysis of the average spectra of local type 1 and type 2
AGN observed with BeppoSAX; yet \citet{burlon11} have recently
obtained the opposite trend from the analysis of the type 1 and type 2
AGN observed in the SWIFTBAT, although with large
uncertainties. Further detailed studies on this particular aspect are
clearly needed.

 \section{Conclusions}
We have analyzed the X-ray spectra corresponding to all identified AGN
within the XBS sample.  \\ From the individual analysis and according
to our fitting criteria, we find that

\begin{itemize}
\item Most AGN are well fitted by a simple unabsorbed power law
  model. The most common deviation from this shape are neutral
  intrinsic absorption and soft-excess emission.
\item In agreement with the AGN unified model, most type 2 AGN are
  absorbed (N$_{H}$ $>$ 4$\times$10$^{21}$ cm$^{-2}$) and by larger
  amounts of intrinsic material than type 1 AGN, which are most
  unabsorbed. Nonetheless, deviations from this simple version of the
  unified model are found, and are more frequent in type 2 AGN.
\item We find that the fraction of exceptions to the unified model is
  of 5\% for the whole sample (only 3\% if we take into account the
  errors on the measured intrinsic absorption). The fraction of type 1
  AGN among absorbed sources is 17\% and 31\% for the HBSS and the
  BSS, respectively. The different values for the two samples are
  likely due to the larger number of unidentified sources within the
  BSS. The fraction of type 2 AGN among unabsorbed sources turns to be
  3\% and 1\% for the HBSS and the BSS, respectively.
\item We find that the X-ray spectral photon index for type 1 AGN is
  anti-correlated with the hard X-ray luminosity. When the type 1 AGN
  sample is split into high- and low-luminosity subsamples, we find
  that the intrinsic photon index for both samples is different almost
  at the 3$\sigma$ level in the plane photon index vs. intrinsic
  dispersion. We compute an average photon index and intrinsic
  dispersion of $\langle\Gamma\rangle$=2.11$\pm$0.04 (2.00$\pm$0.05)
  and $\sigma$=0.29$\pm$0.04 (0.20$\pm$0.04) for the low- (high)
  luminosity sample.
\item We find that the so-called ``soft-excess'' is a common
  characteristic of AGN and it clearly displays different properties
  and origin for unabsorbed and absorbed AGN.
\end{itemize}

From the constructed average spectra we find that

\begin{itemize}
\item The average spectrum for type 2 AGN is different in the HBSS and
  the BSS samples as a result of a larger amount of absorbed sources
  in the HBSS. We do not find any significant differences between
  the type 1 AGN average spectra for these two samples.
\item Apart from the amount of absorption, the differences between
  average spectra corresponding to absorbed and unabsorbed AGN are
  caused by an increase in the amount of reflection.
\item We do not detect a significant relativistic Fe K$\alpha$
  emission line on the average spectrum for unabsorbed sources. We
  estimate an upper limit for a broad relativistic contribution to the
  line of 230 eV at the 3$\sigma$ confidence level.
\item When dividing the type 1 AGN sample into high- and
  low-luminosity sources, we find that the narrow Fe K$\alpha$ line EW
  seems to decrease as the luminosity increases, which is consistent
  with the so-called Iwasawa-Taniguchi effect, although the resulting
  values for the high- and low-luminosity subsamples are consistent
  within errors (EW=110$\pm$30 and 80$\pm$30 eV for the low- and
  high-luminosity subsamples, respectively). We find moreover that
  the amount of reflection may also decrease with luminosity, which
  supports models in which the covering fraction of the putative torus
  decreases as the intrinsic luminosity increases.
\end{itemize}

\begin{acknowledgements}
We would like to thank the referee for providing us with constructive
comments and suggestions. We acknowledge financial support from ASI
(grant n.I/088/06/0 and COFIS contract. It is a pleasure to thank
Tommaso Maccacaro, Mike Watson and Valentina Braito for their initial
efforts and contributions to the XBS project.
\end{acknowledgements}

\begin{landscape}
\begin{table}
\caption{Black body model fit results.}
\label{bbfit}      
\begin{tabular}{c c c c c c c c c c c}   
\hline\hline             
Source & Type & z &  $\Gamma$ & N$_{H}$   & kT & f$_{2-10 keV}$ & Log L$_{2-10 keV}$ & $\chi^{2}$/d.o.f & Probability & Sample \\ 
       &      &          &    & 10$^{22}$ &                   & 10$^{-13}$    &                    &                 &             &         \\ 
       &      &          &    & cm$^{-2}$ & keV             & erg cm$^{-2}$ s$^{-1}$ & erg s$^{-1}$ & & (\%) &  \\ 
(1) & (2) & (3) &  (4) & (5) & (6) & (7) & (8) & (9) & (10) & (11)\\ 
\hline
{\bf XBSJ000532.7+200716} & {\bf AGN1} & {\bf 0.119} & {\bf 2.31$^{+0.38}_{-0.33}$} & {\bf 0.02$^{+0.22}_{-0.02}$} & {\bf 0.10$^{+0.01}_{-0.01}$} & {\bf 0.60} & {\bf 42.42} & {\bf 36/35} & {\bf 40.7} & {\bf BSS}\\
{\bf XBSJ005031.1--520012} & {\bf AGN1} & {\bf 0.463} & {\bf 2.03$^{+0.35}_{-0.19}$} & {\bf $<$0.26} & {\bf 0.11$^{+0.04}_{-0.05}$} & {\bf 0.94} & {\bf 43.95} & {\bf 88/78} & {\bf 19.7} & {\bf BSS}\\
{\bf XBSJ015957.5+003309} & {\bf AGN1} & {\bf 0.310} & {\bf 2.01$^{+0.21}_{-0.11}$} & {\bf $<$0.15} & {\bf 0.07$^{+0.03}_{-0.03}$} & {\bf 3.18} & {\bf 44.06} & {\bf 93/80} & {\bf 15.3} & {\bf HBSS,BSS}\\
{\bf XBSJ021808.3--045845} & {\bf AGN1} & {\bf 0.712} & {\bf 1.91$^{+0.07}_{-0.05}$} & {\bf $<$0.03} & {\bf 0.20$^{+0.01}_{-0.02}$} & {\bf 2.58} & {\bf 44.81} & {\bf 446/417} & {\bf 15.9} & {\bf HBSS,BSS}\\
{\bf XBSJ023530.2--523045} & {\bf AGN1} & {\bf 0.429} & {\bf 1.9$^{f}$} & {\bf $<$0.125} & {\bf 0.059$^{+0.021}_{-0.053}$} & {\bf 1.10} & {\bf 43.91} & {\bf 22/18} & {\bf 23.1} & {\bf BSS}\\
{\bf XBSJ023713.5--522734} & {\bf AGN1} & {\bf 0.193} & {\bf 1.92$^{+0.12}_{-0.14}$} & {\bf $<$0.06} & {\bf 0.11$^{+0.01}_{-0.01}$} & {\bf 2.71} & {\bf 43.52} & {\bf 71/87} & {\bf 88.9} & {\bf HBSS,BSS}\\
XBSJ030641.0--283559 & AGN1 & 0.367 & 2.10$^{+0.39}_{-0.47}$ & $<$0.14 & 0.096$^{+0.054}_{-0.058}$ & 0.51 & 43.45 & 33/24 & 11.0 & BSS\\
XBSJ031311.7--765428 & AGN1 & 1.274 & 1.78$^{+0.17}_{-0.16}$ & $<$0.66 & 0.156$^{+0.122}_{-0.077}$ & 0.98 & 44.95 & 59/50 & 18.8 & BSS\\
{\bf XBSJ031851.9--441815} & {\bf AGN1} & {\bf 1.360} & {\bf 1.58$^{+0.49}_{-0.26}$} & {\bf 0.13$^{+1.51}_{-0.13}$} & {\bf 0.13$^{+0.07}_{-0.05}$} & {\bf 0.94} & {\bf 44.93} & {\bf 17/15} & {\bf 32.8} & {\bf BSS}\\
XBSJ052108.5--251913 & AGN1 & 1.196 & 1.70$^{+0.10}_{-0.20}$ & $<$0.28 & 0.27$^{+0.05}_{-0.08}$ & 2.51 & 45.28 & 78/64 & 11.8 & HBSS,BSS\\
{\bf XBSJ065839.5--560813} & {\bf AGN1} & {\bf 0.211} & {\bf 1.86$^{+0.24}_{-0.24}$} & {\bf 0.04$^{+0.08}_{-0.04}$} & {\bf 0.10$^{+0.01}_{-0.01}$} & {\bf 0.71} & {\bf 43.02} & {\bf 66/68} & {\bf 54.1} & {\bf BSS}\\
{\bf XBSJ074312.1+742937} & {\bf AGN1} & {\bf 0.312} & {\bf 1.98$^{+0.07}_{-0.07}$} & {\bf $<$0.01} & {\bf 0.12$^{+0.01}_{-0.01}$} & {\bf 9.79} & {\bf 44.55} & {\bf 190/192} & {\bf 51.7} & {\bf HBSS,BSS}\\
XBSJ095218.9--013643$^{p}$ & AGN1 & 0.01993 & 1.90$^{f}$ & 4.2$^{+4.4}_{-1.2}$ & 0.15$^{+0.01}_{-0.01}$ & 5.73 & 41.92 & 133/57 & 5e-6 & HBSS,BSS\\
XBSJ100926.5+533426 & AGN1 & 1.718 & 1.72$^{+0.20}_{-0.27}$ & 0.17$^{+0.60}_{-0.17}$ & 0.23$^{+0.10}_{-0.07}$ & 0.87 & 45.19 & 41/37 & 30.7 & BSS\\
XBSJ101922.6+412049 & AGN1 & 0.239 & 1.51$^{+0.15}_{-0.11}$ & $<$0.04 & 0.15$^{+0.02}_{-0.02}$ & 2.87 & 43.71 & 107/110 & 57.8 & HBSS,BSS\\
XBSJ140100.0--110942 & AGN1$^{e}$ & 0.164 & 2.31$^{+0.35}_{-0.23}$ & 0.12$^{+0.17}_{-0.03}$ & 0.07$^{+0.01}_{-0.01}$ & 0.45 & 42.61 & 62/61 & 46.1 & BSS\\
{\bf XBSJ141736.3+523028} & {\bf AGN1} & {\bf 0.985} & {\bf 1.74$^{+0.21}_{-0.16}$} & {\bf $<$0.29} & {\bf 0.22$^{+0.05}_{-0.08}$} & {\bf 0.82} & {\bf 44.61} & {\bf 111/97} & {\bf 17.0} & {\bf BSS}\\
{\bf XBSJ153456.1+013033$^{p}$} & {\bf AGN1} & {\bf 0.310} & {\bf 2.27$^{+0.42}_{-0.24}$} & {\bf $<$0.30} & {\bf 0.09$^{+0.04}_{-0.05}$} & {\bf 1.13} & {\bf 43.65} & {\bf 77/62} & {\bf 9.7} & {\bf BSS}\\
XBSJ223547.9--255836 & AGN1 & 0.304 & 1.43$^{+0.23}_{-0.34}$ & $<$0.03 & 0.19$^{+0.03}_{-0.04}$ & 0.91 & 43.43 & 32/32 & 49.5 & BSS\\
{\bf XBSJ225118.0--175951} & {\bf AGN1} & {\bf 0.172} & {\bf 2.09$^{+0.27}_{-0.20}$} & {\bf 0.16$^{+0.17}_{-0.09}$} & {\bf 0.09$^{+0.01}_{-0.01}$} & {\bf 1.25} & {\bf 43.09} & {\bf 95/88} & {\bf 28.9} & {\bf BSS}\\
{\bf XBSJ231601.7--424038} & {\bf AGN1} & {\bf 0.383} & {\bf 1.74$^{+0.42}_{-0.43}$} & {\bf $<$0.10} & {\bf 0.14$^{+0.03}_{-0.03}$} & {\bf 0.75} & {\bf 43.61} & {\bf 56/51} & {\bf 29.0} & {\bf BSS}\\
\hline
\hline
\end{tabular}
\begin{list}{}{}
\item Columns: (1) Source name; (2) Class; (3) Redshift; (4) Photon index; (5) Intrinsic column density; (6) Black body temperature; (7) Observed flux in the 2-10 keV band, de-absorbed by our Galaxy; (8) Intrinsic luminosity in the 2-10 keV band; (9) $\chi^{2}$ to number of degrees of freedom; (10) Null hypothesis probability; (11) Sample the source belongs to.
\item $^{p}$: Null hypothesis probability $<$ 10\%.$^{e}$: Elusive
  AGN. $^{f}$: Fixed parameter. Source and parameters in bold face
  indicates that this model is considered as our best-fit model.
  Note: Errors and upper limits are at 90\% confidence level. Fluxes
  and luminosities refer to the MOS2 calibration,
\end{list}
\end{table}
\end{landscape}
\begin{landscape}
\begin{table}
\caption{Ionized absorbed power law fit results.}
\label{waplfit}      
\begin{tabular}{c c c c c c c c c c c c}   
\hline\hline             
Source & Type & z &  $\Gamma$ & N$_{H}$   & N$_{Hi}$ & $\xi$ & f$_{2-10 keV}$ & Log L$_{2-10 keV}$ & $\chi^{2}$/d.o.f & Probability & Sample \\ 
       &      &          &    & 10$^{22}$ & 10$^{22}$       &           & 10$^{-13}$    &                    &                 &             &         \\ 
       &      &          &    & cm$^{-2}$ & cm$^{-2}$ &           & erg cm$^{-2}$ s$^{-1}$ & erg s$^{-1}$ & & (\%) &  \\ 
(1) & (2) & (3) &  (4) & (5) & (6) & (7) & (8) & (9) & (10) & (11) & (12) \\ 
\hline
XBSJ005031.1--520012  & AGN1 & 0.463 & 2.30$^{+0.23}_{-0.11}$ & $<$0.049 & 0.60$^{+1.73}_{-0.45}$ & 300$^{+2300}_{-250}$ & 0.79 & 43.89 & 91/78 & 14.8 & BSS\\
XBSJ015957.5+003309$^{p}$  & AGN1 & 0.310 & 2.30$^{+0.14}_{-0.12}$ & $<$0.025 & 0.69$^{+4.67}_{-0.29}$ & 160$^{+2100}_{-110}$ & 2.79 & 44.03 & 92/80 & 8.1 & HBSS,BSS\\
XBSJ023713.5--522734 & AGN1 & 0.193 & 2.34$^{+0.08}_{-0.09}$ & $<$0.016 & 2.34$^{+0.83}_{-1.01}$ & 900$^{+4100}_{-500}$ & 2.35 & 43.52 & 80/87 & 69.3 & HBSS,BSS\\
{\bf XBSJ030641.0--283559} & {\bf AGN1} & {\bf 0.367} & {\bf 2.31$^{+0.23}_{-0.11}$} & {\bf $<$0.06} & {\bf 11.90$^{+11.40}_{-7.14}$} & {\bf 3884$^{+1116}_{-1744}$} & {\bf 0.57} & {\bf 43.60} & {\bf 27/24} & {\bf 29.2} & {\bf BSS}\\
{\bf XBSJ052543.6--334856} & {\bf AGN1} & {\bf 0.735} & {\bf 2.44$^{+0.43}_{-0.45}$} & {\bf 0.14$^{+0.17}_{-0.14}$} & {\bf 6.46$^{+5.14}_{-4.33}$} & {\bf 1561$^{+660}_{-1270}$} & {\bf 0.47} & {\bf 44.27} & {\bf 19/17} & {\bf 34.8} & {\bf BSS}\\
XBSJ065839.5--560813 & AGN1 & 0.211 & 2.85$^{+0.18}_{-0.13}$ & $<$0.016 & 5.0$^{+2.0}_{-2.0}$ & 1600$^{+400}_{-300}$ & 0.52 & 42.94 & 76/67 & 21.6 & BSS\\
XBSJ074312.1+742937 & AGN1 & 0.312 & 2.23$^{+0.04}_{-0.03}$ & $<$0.006 & 3.2$^{+1.3}_{-1.3}$ & 1300$^{+500}_{-500}$ & 9.13 & 44.58 & 206/192 & 23.22 & HBSS,BSS\\
XBSJ101922.6+412049 & AGN1 & 0.239 & 1.82$^{+0.09}_{-0.04}$ & $<$0.02 & 8.3$^{+9.5}_{-3.3}$ & 1700$^{+1600}_{-800}$ & 2.72 & 43.70 & 109/110 & 71.1 & HBSS,BSS\\
{\bf XBSJ140100.0--110942} & {\bf AGN1} & {\bf 0.164} & {\bf 2.52$^{+0.28}_{-0.11}$} & {\bf $<$0.03} & {\bf 0.43$^{+0.35}_{-0.18}$} & {\bf 60$^{+130}_{-50}$} & {\bf 0.41} & {\bf 42.60} & {\bf 63/61} & {\bf 41.1} & {\bf BSS}\\
{\bf XBSJ140127.7+025605} & {\bf AGN1} & {\bf 0.265} & {\bf 1.84$^{+0.10}_{-0.05}$} & {\bf 0.17$^{+0.04}_{-0.04}$} & {\bf 0.60$^{+0.38}_{-0.27}$} & {\bf 38$^{+51}_{-30}$} & {\bf 6.81} & {\bf 44.22} & {\bf 381/359} & {\bf 22.2} & {\bf HBSS,BSS}\\
{\bf XBSJ223547.9--255836} & {\bf AGN1} & {\bf 0.304} & {\bf 2.11$^{+0.27}_{-0.25}$} & {\bf 0.08$^{+0.09}_{-0.08}$} & {\bf 15.60$^{+14.20}_{-8.64}$} & {\bf 2772$^{+2061}_{-1321}$} & {\bf 0.78} & {\bf 43.41} & {\bf 31/32} & {\bf 50.7} & {\bf BSS}\\
XBSJ225118.0--175951 & AGN1 & 0.1718 & 2.89$^{+0.11}_{-0.24}$ & 0.05$^{+0.02}_{-0.01}$ & 4.3$^{+1.5}_{-1.1}$ & 4900$^{+100}_{-2700}$ & 1.07 & 43.14 & 104/88 & 12.4 & BSS\\
\hline
\hline
\end{tabular}
\begin{list}{}{}
\item Columns: (1) Source name; (2) Class; (3) Redshift; (4) Photon index; (5) Intrinsic column density; (6) Ionized absorber column density; (7) Ionization parameter; (8) Observed flux in the 2-10 keV band, de-absorbed by our Galaxy; (9) Intrinsic luminosity in the 2-10 keV band; (10) $\chi^{2}$ to number of degrees of freedom; (11) Null hypothesis probability; (12) Sample the source belongs to.
\item $^{p}$: Null hypothesis probability $<$ 10\%.$^{e}$: Elusive AGN. $^{f}$: Fixed parameter. Source and parameters in bold face indicates that this model is considered as our best-fit model. Note: Errors and upper limits are at 90\% confidence level. Fluxes and luminosities refer to the MOS2 calibration.
\end{list}
\end{table}
\begin{table}
\caption{Power law plus reflected component fit results.}
\label{plreffit}      
\begin{tabular}{c c c c c c c c c c c}   
\hline\hline             
Source & Type & z & $\Gamma$ & N$_{H}$   & R & f$_{2-10 keV}$ & Log L$_{2-10 keV}$ & $\chi^{2}$/d.o.f & Probability & Sample \\ 
       &      &          &    & 10$^{22}$ &                   & 10$^{-13}$    &                    &                 &             &         \\ 
       &      &          &    & cm$^{-2}$ &             & erg cm$^{-2}$ s$^{-1}$ & erg s$^{-1}$ & & \% &  \\ 
(1) & (2) & (3) &  (4) & (5) & (6) & (7) & (8) & (9) & (10) & (11)\\ 
\hline
XBSJ021808.3--045845$^{p}$ & AGN1 & 0.712 & 2.38$^{+0.19}_{-0.04}$ & $<$0.04 & 0.3($<$0.5) & 2.67 & 44.80 & 459/417 & 7.8 & HBSS,BSS\\
XBSJ023713.5--522734 & AGN1 & 0.193 & 2.80$^{+0.08}_{-0.15}$ & $<$0.017 & 2($<$7) & 2.95 & 43.53 & 75/88 & 84.4 & HBSS,BSS\\
{\bf XBSJ031311.7--765428} & {\bf AGN1} & {\bf 1.274} & {\bf 2.16$^{+0.75}_{-0.16}$} & {\bf $<$0.25} & {\bf $<$0.10} & {\bf 1.11} & {\bf 44.94} & {\bf 60/50} & {\bf 17.2} & {\bf BSS}\\
{\bf XBSJ043448.3--775329} & {\bf AGN1$^{e}$} & {\bf 0.097} & {\bf 1.9$^{f}$} & {\bf 0.20$^{+0.27}_{-0.20}$} & {\bf 0.4($<$0.6)} & {\bf 4.00} & {\bf 43.00} & {\bf 9/11} & {\bf 62.1} & {\bf BSS}\\
{\bf XBSJ052108.5--251913} & {\bf AGN1} & {\bf 1.196} & {\bf 2.22$^{+0.51}_{-0.15}$} & {\bf $<$0.17} & {\bf 0.6($<$1.20)} & {\bf 2.73} & {\bf 45.28} & {\bf 78/64} & {\bf 11.2} & {\bf HBSS,BSS}\\
XBSJ074312.1+742937 & AGN1 & 0.312 & 2.51$^{+0.06}_{-0.06}$ & $<$0.006 & 1.3($<$1.6) & 10.6 & 44.56 & 199/192 & 35.2 & HBSS,BSS\\
{\bf XBSJ101922.6+412049} & {\bf AGN1} & {\bf 0.239} & {\bf 2.12$^{+0.27}_{-0.08}$} & {\bf $<$0.03} & {\bf 0.3($<0.6$} & {\bf 2.97} & {\bf 43.65} & {\bf 102/110} & {\bf 69.0} & {\bf HBSS,BSS}\\
XBSJ141736.3+523028 & AGN1 & 0.985 & 2.16$^{+0.60}_{-0.11}$ & $<$0.14 & 0.3($<$1.7) & 0.89 & 44.60 & 112/97 & 15.0 & BSS\\
XBSJ223547.9--255836 & AGN1 & 0.304 & 2.88$^{+0.68}_{-0.69}$ & 0.08$^{+0.09}_{-0.08}$ & 1.7($<$5) & 0.86 & 43.42 & 32/32 & 47.7 & BSS\\
XBSJ231601.7--424038 & AGN1 & 0.383 & 3.05$^{+0.78}_{-0.25}$ & $<$0.072 & 2($<5.5$) & 0.73 & 43.59 & 56/52 & 33.9 & BSS\\
\hline
\hline
\end{tabular}
\begin{list}{}{}
\item Columns: (1) Source name; (2) Class; (3) Redshift; (4) Photon index; (5) Intrinsic column density; (6) Reflection scaling factor and upper limit at 90\% confidence (between parenthesis); (7) Observed flux in the 2-10 keV band, de-absorbed by our Galaxy; (8) Intrinsic luminosity in the 2-10 keV band; (9) $\chi^{2}$ to number of degrees of freedom; (10) Null hypothesis probability; (11) Sample the source belongs to.
\item $^{p}$: Null hypothesis probability $<$ 10\%.$^{e}$: Elusive AGN. $^{f}$: Fixed parameter. Source and parameters in bold face indicates that this model is considered as our best-fit model. Note: Errors and upper limits are at 90\% confidence level. Fluxes and luminosities refer to the MOS2 calibration.
\end{list}
\end{table}
\end{landscape}
\begin{landscape}
\begin{table}
\caption{Leaky/Leaky+line fit results.}
\label{leakyfit}      
\begin{tabular}{c c c c c c c c c c c c c c}   
\hline\hline             
Source & Type & z & $\Gamma$ & N$_{H}$   & Ratio & E & $\sigma$ & EW & f$_{2-10 keV}$ & Log L$_{2-10 keV}$ & $\chi^{2}$/d.o.f & Probability & Sample \\ 
       &      &          &    & 10$^{22}$ &    & & &              & 10$^{-13}$    &                    &                 &             &         \\ 
       &      &          &    & cm$^{-2}$ &  & keV & eV & eV           & erg cm$^{-2}$ s$^{-1}$ & erg s$^{-1}$ & & \% &  \\ 
(1) & (2) & (3) & (4) & (5) & (6) & (7) & (8) & (9) & (10) & (11) & (12) & (13) & (14) \\ 
\hline
{\bf XBSJ021822.2--050615$^{p}$} & {\bf AGN2$^{e}$} & {\bf 0.044} & {\bf 1.97$^{+0.13}_{-0.24}$} & {\bf 22.20$^{+2.63}_{-1.55}$} & {\bf 0.004$^{+0.003}_{-0.001}$} & {\bf 6.28$^{+0.15}_{-0.10}$} & {\bf 210$^{+120}_{-70}$} & {\bf 370$^{+190}_{-190}$} & {\bf 2.81} & {\bf 42.57} & {\bf 112/79} & {\bf 0.8} & {\bf HBSS}\\
{\bf XBSJ033845.7--352253} & {\bf AGN2} & {\bf 0.113} & {\bf 1.99$^{+0.61}_{-0.40}$} & {\bf 31.30$^{+9.70}_{-8.20}$} & {\bf 0.024$^{+0.072}_{-0.022}$} & {\bf \dots} & {\bf \dots} & {\bf \dots} & {\bf 1.70} & {\bf 43.26} & {\bf 42/36} & {\bf 22.3} & {\bf HBSS}\\
{\bf XBSJ040758.9--712833} & {\bf AGN2} & {\bf 0.134} & {\bf 1.90$^{f}$} & {\bf 21.90$^{+14.10}_{-10.60}$} & {\bf 0.032$^{+0.057}_{-0.030}$} & {\bf \dots} & {\bf \dots} & {\bf \dots} & {\bf 2.23} & {\bf 43.43} & {\bf 11/11} & {\bf 42.7} & {\bf HBSS}\\
{\bf XBSJ091828.4+513931} & {\bf AGN1} & {\bf 0.185} & {\bf 1.85$^{+0.43}_{-0.41}$} & {\bf 6.41$^{+2.17}_{-2.05}$} & {\bf 0.045$^{+0.060}_{-0.029}$} & {\bf \dots} & {\bf \dots} & {\bf \dots} & {\bf 2.56} & {\bf 43.58} & {\bf 12/15} & {\bf 67.8} & {\bf HBSS,BSS}\\
{\bf XBSJ095218.9--013643} & {\bf AGN1} & {\bf 0.020} & {\bf 3.38$^{+0.12}_{-0.04}$} & {\bf 49.00$^{+6.60}_{-5.60}$} & {\bf 0.007$^{+0.002}_{-0.002}$} & {\bf 0.92$^{+0.02}_{-0.03}$} & {\bf 50$^{+30}_{-220}$} & {\bf 150$^{+40}_{-40}$} & {\bf 9.96} & {\bf 43.03} & {\bf 60/54} & {\bf 27.0} & {\bf HBSS,BSS}\\
{\bf XBSJ112026.7+431520} & {\bf AGN2$^{e}$} & {\bf 0.146} & {\bf 1.88$^{+0.66}_{-0.90}$} & {\bf 8.25$^{+2.25}_{-3.56}$} & {\bf 0.021$^{+0.057}_{-0.029}$} & {\bf \dots} & {\bf \dots} & {\bf \dots} & {\bf 1.88} & {\bf 43.26} & {\bf 20/15} & {\bf 15.8} & {\bf HBSS}\\
\hline
\hline
\end{tabular}
\begin{list}{}{}
\item Columns: (1) Source name; (2) Class; (3) Redshift; (4) Photon index; (5) Intrinsic column density; (6) Scattered component to direct component ratio; (7) Emission line central energy; (8) Emission line width; (9) Emission line equivalent width; (10) Observed flux in the 2-10 keV band, de-absorbed by our Galaxy; (11) Intrinsic luminosity in the 2-10 keV band; (12) $\chi^{2}$ to number of degrees of freedom; (13) Null hypothesis probability; (14) Sample the source belongs to.
\item $^{p}$: Null hypothesis probability $<$ 10\%.$^{e}$: Elusive AGN. $^{f}$: Fixed parameter. Source and parameters in bold face indicates that this model is considered as our best-fit model. Note: Errors and upper limits are at 90\% confidence level. Fluxes and luminosities refer to the MOS2 calibration.
\end{list}
\end{table}
\begin{table}
\caption{Power law plus thermal component fit results.}
\label{pltfit}      
\begin{tabular}{c c c c c c c c c c c c c c}   
\hline\hline             
Source & Type & z & $\Gamma$ & N$_{H}$   & kT & E & $\sigma$ & EW & f$_{2-10 keV}$ & Log L$_{2-10 keV}$ & $\chi^{2}$/d.o.f & Probability & Sample \\ 
       &      &          &    & 10$^{22}$ & & &  &                 & 10$^{-13}$    &                    &                 &             &         \\ 
       &      &          &    & cm$^{-2}$ & keV & keV & eV & eV             & erg cm$^{-2}$ s$^{-1}$ & erg s$^{-1}$ & & \% &  \\ 
(1) & (2) & (3) &  (4) & (5) & (6) & (7) & (8) & (9) & (10) & (11) & (12) & (13) & (14)\\ 
\hline
{\bf XBSJ012654.3+191246} & {\bf AGN$^{e}$} & {\bf 0.043} & {\bf 1.90$^{f}$} & {\bf 0.066$^{+0.968}_{-0.066}$} & {\bf 1.04$^{+0.29}_{-0.11}$} & {\bf \dots} & {\bf \dots} & {\bf \dots} & {\bf 0.29} & {\bf 41.17} & {\bf 7/11} & {\bf 83.0} & {\bf BSS}\\
XBSJ021822.2--050615$^{p}$ & AGN2 & 0.044 & 1.67$^{+0.13}_{-0.19}$ & 19.6$^{+1.8}_{-1.7}$ & 1.8$^{+1.6}_{-0.4}$ & 6.28$^{+0.15}_{-0.40}$ & 210$^{+120}_{-70}$ & 360$^{+20}_{-20}$ & 2.93 & 42.51 & 109/78 & 1.2 & HBSS\\
XBSJ033845.7--322253$^{p}$ & AGN2 & 0.113 & 1.9$^{f}$ & 24.1$^{+8.7}_{-6.5}$ & 0.21$^{+0.5}_{-0.5}$ & \dots & \dots & \dots & 1.61 & 43.17 & 49/36 & 7.3 & HBSS\\
XBSJ040758.9--712833 & AGN2 & 0.134 & 1.9$^{f}$ & 11.0$^{+8.2}_{-4.2}$ & 0.24$^{+0.10}_{-0.08}$ & \dots & \dots & \dots & 1.80 & 43.22 & 14/10 & 16.3 & HBSS\\
XBSJ112026.7+431520 & AGN2 & 0.146 & 1.9$^{f}$ & 7.5$^{+1.7}_{-1.3}$ & 0.87$^{+0.45}_{-0.25}$ & \dots & \dots & \dots & 1.87 & 43.26 & 18/15 & 25.5 & HBSS\\
{\bf XBSJ231546.5--590313} & {\bf AGN2} & {\bf 0.0446} & {\bf 1.90$^{f}$} & {\bf 1.25$^{+0.58}_{-0.51}$} & {\bf 0.46$^{+0.10}_{-0.18}$} & {\bf \dots} & {\bf \dots} & {\bf \dots} & {\bf 1.49} & {\bf 41.95} & {\bf 15/14} & {\bf 35.9} & {\bf BSS}\\
\hline
\hline
\end{tabular}
\begin{list}{}{}
\item Columns: (1) Source name; (2) Class; (3) Redshift; (4) Photon index; (5) Intrinsic column density; (6) Plasma temperature; (7) Emission line central energy; (8) Emission line width; (9) Emission line equivalent width; (10) Observed flux in the 2-10 keV band, de-absorbed by our Galaxy; (11) Intrinsic luminosity in the 2-10 keV band; (12) $\chi^{2}$ to number of degrees of freedom; (13) Null hypothesis probability; (14) Sample the source belongs to.
\item $^{p}$: Null hypothesis probability $<$ 10\%.$^{e}$: Elusive AGN. $^{f}$: Fixed parameter. Source and parameters in bold face indicates that this model is considered as our best-fit model. Note: Errors and upper limits are at 90\% confidence level. Fluxes and luminosities refer to the MOS2 calibration.
\end{list}
\end{table}
\end{landscape}
\begin{landscape}
\begin{table}
\caption{Power law and absorption edges fit results.}
\label{pledgefit}      
\begin{tabular}{c c c c c c c c c c c c}   
\hline\hline             
Source & Type & z &  $\Gamma$ & N$_{H}$   & E & $\tau$ & f$_{2-10 keV}$ & Log L$_{2-10 keV}$ & $\chi^{2}$/d.o.f & Probability & Sample \\ 
       &      &          &    & 10$^{22}$ &     &           & 10$^{-13}$    &                    &                 &             &         \\ 
       &      &          &    & cm$^{-2}$ & keV &             & erg cm$^{-2}$ s$^{-1}$ & erg s$^{-1}$ & & \% &  \\ 
(1) & (2) & (3) &  (4) & (5) & (6) & (7) & (8) & (9) & (10) & (11) & (12) \\ 
\hline
{\bf XBSJ100926.5+533426} & {\bf AGN1} & {\bf 1.718} & {\bf 2.02$^{+0.08}_{-0.11}$} & {\bf $<$0.18} & {\bf 2.09$^{+0.14}_{-0.21}$} & {\bf 0.63$^{+0.37}_{-0.33}$} & {\bf 0.73} & {\bf 45.18} & {\bf 40/37} & {\bf 34.5} & {\bf BSS}\\
{\bf XBSJ102412.3+042023} & {\bf AGN1} & {\bf 1.458} & {\bf 2.01$^{+0.16}_{-0.10}$} & {\bf $<$0.17} &  {\bf 3.25$^{+0.13}_{-0.13}$} &  {\bf 0.9$^{+0.5}_{-0.4}$} & {\bf 0.53} & {\bf 44.87} & {\bf 37/33} & {\bf 28.9} & {\bf BSS}\\
{\bf XBSJ204159.2--321439} & {\bf AGN1} & {\bf 0.738} & {\bf 2.08$^{+0.18}_{-0.11}$} & {\bf $<$0.10} & {\bf 2.10$^{+0.15}_{-0.11}$} & {\bf 0.70$^{+0.30}_{-0.30}$} & {\bf 1.05} & {\bf 44.45} & {\bf 27/26} & {\bf 42.3} & {\bf BSS}\\
\hline
\hline
\end{tabular}
\begin{list}{}{}
\item Columns: (1) Source name; (2) Class; (3) Redshift; (4) Photon index; (5) Intrinsic column density; (6) Threshold energy; (7) Absorption depth at threshold; (8) Observed flux in the 2-10 keV band, de-absorbed by our Galaxy; (9) Intrinsic luminosity in the 2-10 keV band; (10) $\chi^{2}$ to number of degrees of freedom; (11) Null hypothesis probability; (12) Sample the source belongs to.
\item $^{p}$: Null hypothesis probability $<$ 10\%.$^{e}$: Elusive AGN. $^{f}$: Fixed parameter. Source and parameters in bold face indicates that this model is considered as our best-fit model. Note: Errors and upper limits are at 90\% confidence level. Fluxes and luminosities refer to the MOS2 calibration.
\end{list}
\end{table}
\end{landscape}

\bibliographystyle{aa} 
\bibliography{15227refs} 

\begin{appendix}
\section{Notes on individual sources}
\begin{figure*}[ht]
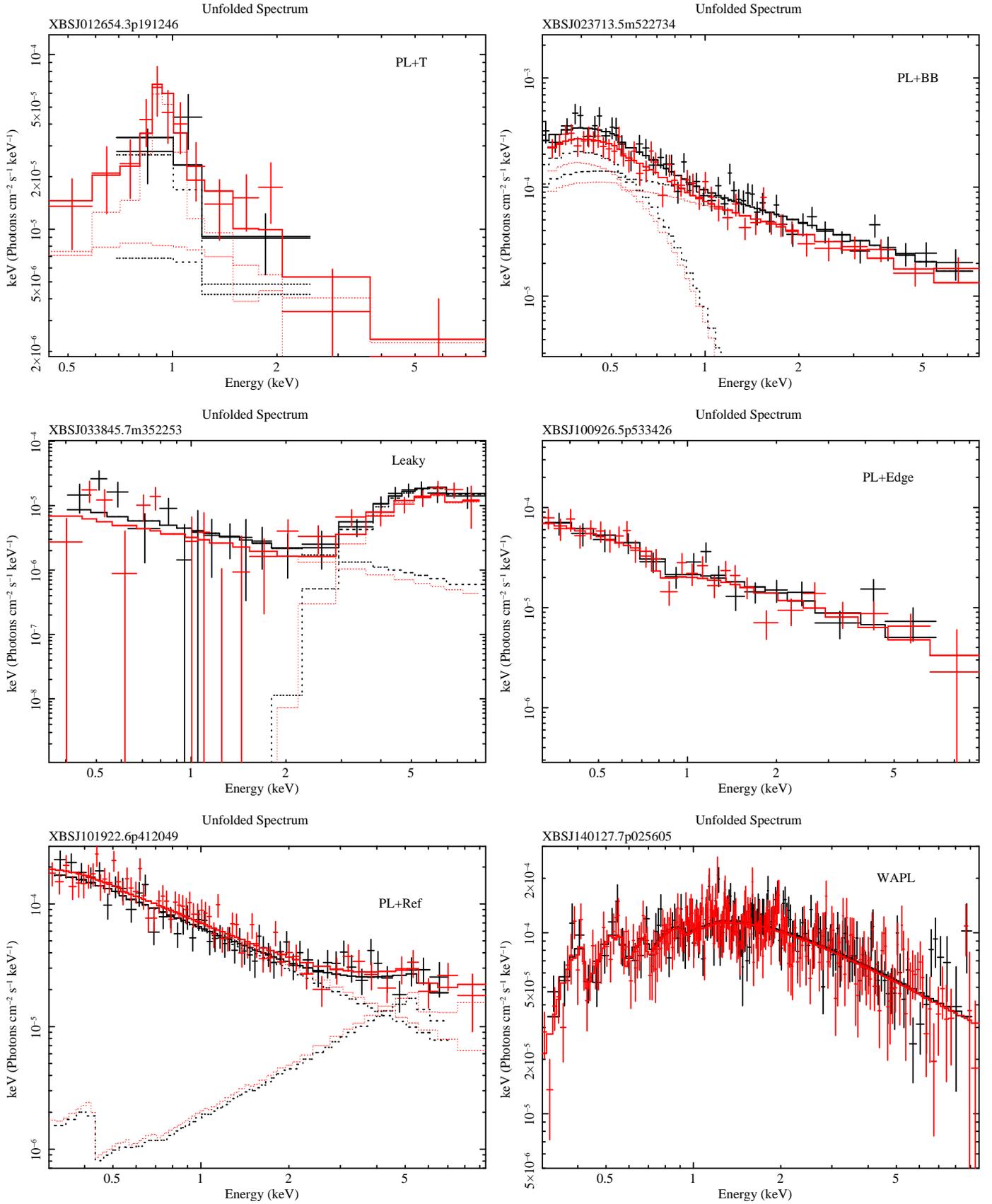

  \centering
$$
\begin{array}{cc}
    \includegraphics[angle=-90,width=9cm]{15227fga1a.ps} & \includegraphics[angle=-90,width=9cm]{15227fga1b.ps}\\
\includegraphics[angle=-90,width=9cm]{15227fga1c.ps} & \includegraphics[angle=-90,width=9cm]{15227fga1d.ps}\\
\includegraphics[angle=-90,width=9cm]{15227fga1e.ps} & \includegraphics[angle=-90,width=9cm]{15227fga1f.ps}\\
  \end{array}
$$ 
 \caption{Unfolded spectra corresponding to the different additional components used during the spectral fit. From top to bottom and left to right: Power law plus thermal component, power law plus black body, leaky model, power law and absorption edge, power law plus reflection component and power law and ionized absorber.}
 \label{examples}
\end{figure*}

During the spectral fit, those sources which are not well fitted using
a simple power law model can be usually well fitted by using different
additional components. Here, we describe how we decided between the
different models that are an acceptable fit for each source: \\
\begin{itemize}
\item {\bf Leaky model:} All sources for which a leaky model was
  selected as our best-fit model share a common spectral shape: a
  power law shape at high energies that drops around $\sim$2-3 keV and
  an additional soft component. These sources are four type 2 AGN
  (XBSJ021822.2--050615, XBSJ033845.7--322253, XBSJ040758.9--712833
  and XBSJ112026.7+431520) and two type 1 AGN (XBSJ091828.4+513931 and
  XBSJ095218.9--013643). XBSJ095218.9--013643 is a NLSy1 (narrow line
  Seyfert 1) whose intriguing X-ray spectral shape (a very steep
  photon index and large amount of absorption that partially covers
  the central source) and variability (variability of a factor of 4 in
  the soft X-rays) have been already studied in detail and presented
  in \citet{grupe04}. In Fig.~\ref{examples} is shown an example of a
  leaky model fit. We find that leaving the soft photon index free to
  vary for all these sources does not significantly improve the
  fit. However, in all cases, this soft photon index steepens if it is
  left free to vary, which suggests the contribution of an additional
  soft component, most likely a thermal component given the low
  luminosity observed for this sources. In no case adding a thermal
  component to the leaky model significantly improves the fit, and by
  fitting a simple absorbed power law plus a thermal component always
  gives worse residuals at low energies than the leaky model.  \\
\item {\bf Warm absorption:} We find that an additional ionized
  absorber gives a best fit in five cases (XBSJ030641.0--283559,
  XBSJ052543.6--334856, XBSJ140100.0--110942, XBSJ140127.7+025605,
  XBSJ223547.9--255836, all type 1 AGN), although the ionized absorber
  parameters, mainly the ionization state of the absorber, are not
  well constrained in all cases. We selected this model as our
  best-fit model when the power-law residuals at low energies showed
  some evidence of an structured shape resembling absorption lines or
  edges. In two cases, the source also displays a soft-excess
  (XBSJ030641.0--283559 and XBSJ223547.9--255836). One example of this
  model is again shown in Fig.~\ref{examples}. The ionized absorber
  was added to the neutral one because of the way the spectral fit is
  carried out, i.e., our base-line model is a simple power law
  including neutral intrinsic absorption. It is worth noting, however,
  that none of the sources for which the best-fit model includes warm
  absorption need significant additional cold absorption, as can be
  seen in Table~\ref{waplfit}. \\
\item {\bf Absorption edges:} In three cases (XBSJ100926.5+533426,
  XBSJ102412.3+042023 and XBSJ204159.2--321439), an absorption edge
  has to be added to the simple power law model to obtain an
  acceptable fit. It is not clear whether these edges are real or an
  instrumental effect given the energies at which they are found, but
  they could be caused by a warm absorber that our simple {\tt Xspec
    absori} model is not able to fit properly. See again
  Fig.~\ref{examples}.  \\
\item {\bf Reflection component:} We find that a simple power law plus
  a neutral reflection component is a good fit in four cases
  (XBSJ031311.7--765428, XBSJ043448.3--775329, XBSJ052108.5--251913
  and XBSJ101922.6+412049; all type 1 AGN). We use neutral reflection
  in all cases ({\tt pexrav} model in {\tt Xspec}) since our intention
  is not to determine where this reflection component originates
  in. Given the data quality, we can only estimate the amount of
  reflection by the reflection fraction R in the {\tt pexrav}
  model. Nevertheless, we find that in all cases but one
  (XBSJ031311.7--765428) the reflection component is most likely
  coming from Compton-thick material far away from the central source,
  the putative torus in unified models, given the spectral shape, a
  rather flat continuum at high energies, and the characteristics of a
  possible Fe K$\alpha$ line. Note also that all but
  XBSJ043448.3-775329 show a soft-excess.  \\
\item {\bf Black body:} A phenomenological black body component is
  needed to obtain an acceptable fit in 13 cases (XBSJ000532.7+200716,
  XBSJ005031.1--520012, XBSJ015957.5+003309, XBSJ021808.3--045845,
  XBSJ023530.2--523045, XBSJ023713.5--522734, XBSJ031851.9--441815,
  XBSJ065839.5--560813, XBSJ074312.1+742937, XBSJ141736.3+523028,
  XBSJ153456.1+013033, XBSJ225118.0--175951 and XBSJ231601.7--424038;
  all type 1 AGN), all showing soft-excess. The physical origin for
  this soft component is not clear although a host galaxy thermal
  contribution is ruled out given its high luminosity in all cases but
  XBSJ000532.7+200716. In that case the low luminosity found for the
  black body component ($\sim$4$\times$10$^{42}$ erg s$^{-1}$ in the
  0.5-2.0 keV energy range) could be caused by thermal emission, but
  adding an {\tt Xspec mekal} component does not improve the simple
  power law fit. More complex models recently proposed in the
  literature \citep{crummy06,midd07} cannot be used in our case given
  the data quality, and in any case, they are indistinguishable in the
  EPIC-covered energy range. In some cases, the need for a black body
  component instead of a more physically motivated model, could be
  just due to the data quality. For example, in seven cases,
  XBSJ005031.1--520012, XBSJ015957.5+003309, XBSJ021808.3--045845,
  XBSJ065839.5--560813, XBSJ074312.1+742937, XBSJ153456.1+013033 and
  XBSJ225118.0--175951 (see Table \ref{waplfit}), an ionized absorber
  is also a good fit, but gives worse residuals that the black body
  model. This could be because of to both the data quality and the
  need of a better representation of the ionized absorber. And for
  XBSJ021808.3-045845, XBSJ023713.5--522734, XBSJ074312.1+742937,
  XBSJ141736.3+523028 and XBSJ231601.7--424038 (see Table
  \ref{plreffit}, note that XBSJ021828.3--045845 and
  XBSJ074312.1+742937 can be also fitted by using ionized absorption),
  the addition of a reflection component instead of a black body also
  significantly improves the fit. For the first two cases, this
  reflection component could derive from ionized material.  \\
\item {\bf Sources for which no best fit is found:} We are unable to
  find an acceptable fit in 11 cases. We do not find that these
  sources share any common characteristic, and that an acceptable fit
  is not found could be simply due to our selection criteria based on
  the resulting null hypothesis probability. The simple power law fit
  for these sources is shown in Fig.~\ref{nbff}. In the case of
  XBSJ021822.2--050615 and XBSJ153456.1+013033, the fit corresponds to
  a leaky model and a power law plus a black body, which significantly
  improve the fit, but not enough to obtain a probability $>$ 10\%.
\end{itemize}

\begin{figure*}[ht]
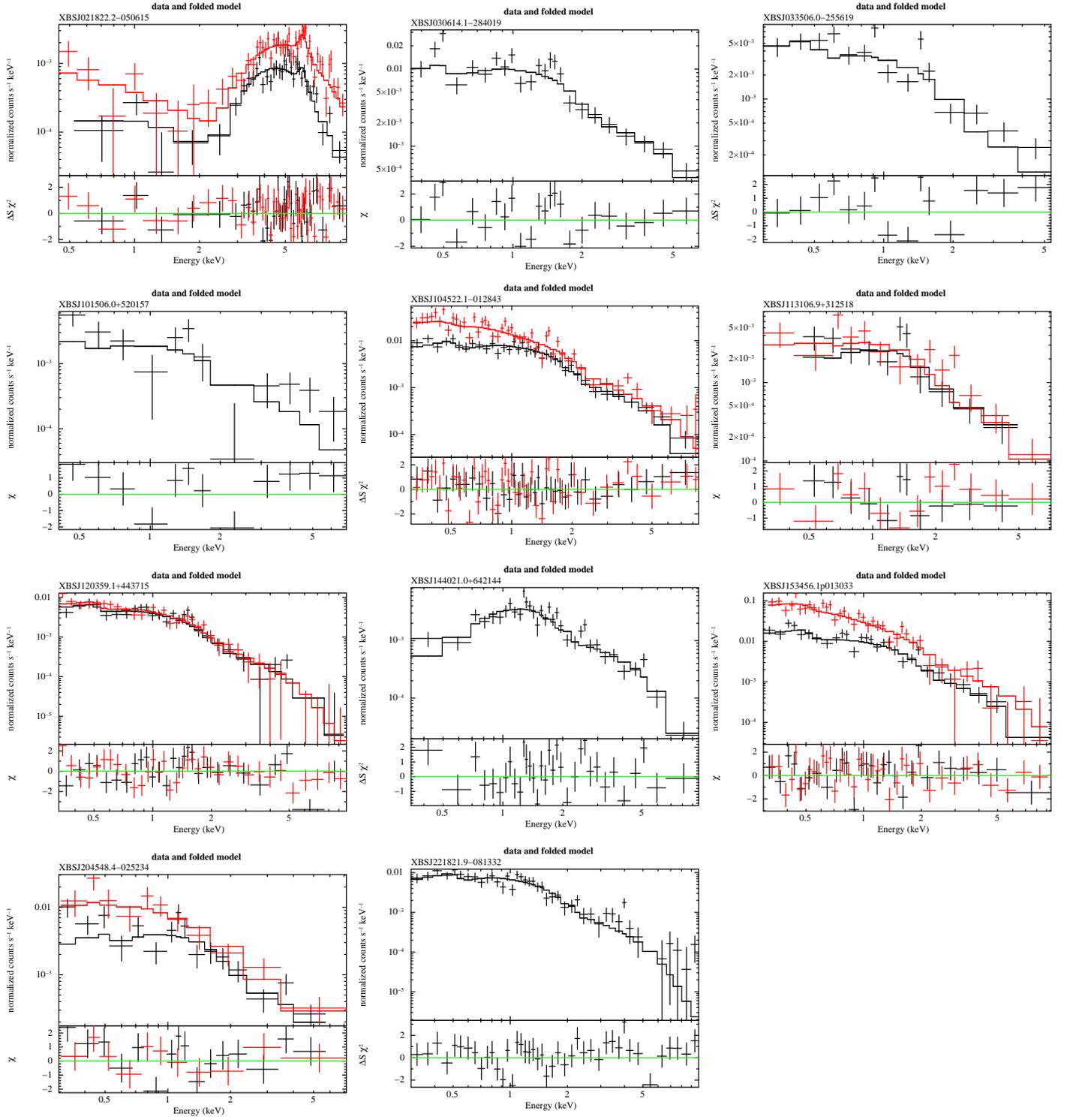

  \centering
$$
  \begin{array}{ccc}
    \includegraphics[angle=-90,width=6cm]{15227fga2a.ps} & \includegraphics[angle=-90,width=6cm]{15227fga2b.ps} & \includegraphics[angle=-90,width=6cm]{15227fga2c.ps}\\
\includegraphics[angle=-90,width=6cm]{15227fga2d.ps} & \includegraphics[angle=-90,width=6cm]{15227fga2e.ps} & \includegraphics[angle=-90,width=6cm]{15227fga2f.ps}\\
\includegraphics[angle=-90,width=6cm]{15227fga2g.ps} & \includegraphics[angle=-90,width=6cm]{15227fga2h.ps} & \includegraphics[angle=-90,width=6cm]{15227fga2i.ps} \\
\includegraphics[angle=-90,width=6cm]{15227fga2j.ps} & \includegraphics[angle=-90,width=6cm]{15227fga2k.ps} & \\
  \end{array}
$$ 
 \caption{Data and residuals corresponding to the eleven sources for which no acceptable fit is found.}
 \label{nbff}
\end{figure*}

\end{appendix}
\Online
\onllongtab{2}{
\begin{landscape}
\begin{longtable}{lcccccccccc}
\caption{\label{obsids} Summary of {\it XMM-Newton} observations used.}\\
\hline\hline
Source name & OBSID & & FILTER & & N$_{H}$Gal &  & EXPOSURE TIME &  & TOTAL COUNTS & SAMPLE \\
 &  & pn & MOS1 & MOS2 & & pn & MOS1  &  MOS2 & (0.3-10 keV) & \\
 &  & & & & (10$^{20}$cm$^{-2}$) & (ks) & (ks)  &  (ks) &  & \\
 (1) & (2) & (3) & (4) & (5) & (6)  &  (7) & (8) & (9) & (10) & (11) \\
\hline
\endfirsthead
\caption{continued.}\\
\hline\hline
Source name & OBSID & & FILTER & & N$_{H}$Gal &  & EXPOSURE TIME &  & TOTAL COUNTS & SAMPLE \\
 &  & pn & MOS1 & MOS2 & & pn & MOS1  &  MOS2 & (0.3-10 keV) \\
 &  & & & & (10$^{20}$cm$^{-2}$) & (ks) & (ks)  &  (ks) &  \\
 (1) & (2) & (3) & (4) & (5) & (6)  &  (7) & (8) & (9) & (10) & (11) \\
\hline
\endhead
\hline
\endfoot
XBSJ000027.7--250442 & 0125310101 & Med & Med & Med & 1.88 & 21.28 & 18.62 & 17.98 & 1270 & BSS\\
XBSJ000031.7--245502 & 0125310101 & Med & Med & Med & 1.88 & 21.28 & 18.65 & 17.98 & 705 & BSS\\
XBSJ000100.2--250501 & 0125310101 & Med & Med & Med & 1.88 & 21.27 & 18.64 & 17.98 & 830 & BSS\\
XBSJ000102.4--245850 & 0125310101 & Med & Med & Med & 1.88 & 21.28 & 18.65 & 17.98 & 810 & BSS\\
XBSJ000532.7+200716 & 0101040101 & \dots & Med & Med & 3.99 & \dots & 31.38 & 30.71 & 1000 & BSS\\
XBSJ001831.6+162925 & 0111000101 & Med & Med & Med & 4.07 & 22.75 & 29.62 & 29.33 & 2950 & BSS\\
XBSJ002618.5+105019 & 0001930101 & Thin & Thin & Thin & 5.07 & 14.18 & 17.91 & 17.51 & 2348 & HBSS,BSS\\
XBSJ002637.4+165953 & 0050140201 & Thin & Thin & Thin & 4.23 & 37.41 & 46.02 & 45.91 & 3200 & BSS\\
\hline                 
\end{longtable}
\begin{list}{}{}
\item Columns: (1) Source name; (2) Observation identifier; (3), (4), (5) Filter in use during the observation for the pn, MOS1 and MOS2 cameras, respectively; (6) Galactic column density toward the used {\it XMM-Newton} pointing. (7), (8), (9) Resulting exposure time after removing high-background intervals for the pn, MOS1 and MOS2 cameras, respectively; (10) Total EPIC counts (pn+MOS1+MOS2); (11) Sample the source belongs to.
\end{list}
\end{landscape}
}
\onllongtab{3}{
\begin{landscape}
\begin{longtable}{lcccccccccc}
\caption{\label{splfit} Power law fit results.}\\
\hline\hline
Source & Type & z & $\Gamma$ & N$_{H}$ & f$_{2-10 keV}$ & Log L$_{2-10 keV}$ & $\chi^{2}$/d.o.f & Probability & Sample & Best\\ 
 &  &  & & 10$^{22}$ & 10$^{-13}$ & & &  &  & fit\\ 
 &  &  & & cm$^{-2}$ & erg cm$^{-2}$ s$^{-1}$ & erg s$^{-1}$ & & \% & &  \\ 
 (1) & (2) & (3) & (4) & (5) & (6)  &  (7) & (8) & (9) & (10) & (11)\\
\hline
\endfirsthead
\caption{continued.}\\
\hline\hline
Source & Type & z & $\Gamma$ & N$_{H}$ & f$_{2-10 keV}$ & Log L$_{2-10 keV}$ & $\chi^{2}$/d.o.f & Probability & Sample & Best\\ 
 &  &  & & 10$^{22}$ & 10$^{-13}$ & & &  & & fit \\ 
 &  &  & & cm$^{-2}$ & erg cm$^{-2}$ s$^{-1}$ & erg s$^{-1}$ & & \% & &  \\ 
 (1) & (2) & (3) & (4) & (5) & (6)  &  (7) & (8) & (9) & (10) & (11)\\
\hline
\endhead
\hline
\endfoot
XBSJ000027.7--250442 & AGN1 & 0.336 & 1.87$^{+0.10}_{-0.08}$ & $<$0.02 & 0.72 & 43.48 & 64/66 & 55.4 & BSS & Y \\
XBSJ000031.7--245502 & AGN1 & 0.284 & 2.29$^{+0.14}_{-0.13}$ & $<$0.02 & 0.52 & 43.22 & 31/33 & 56.0 & BSS & Y \\
XBSJ000100.2--250501 & AGN1 & 0.850 & 1.67$^{+0.14}_{-0.18}$ & 0.71$^{+0.22}_{-0.24}$ & 1.38 & 44.66 & 48/41 & 22.4 & BSS & Y \\
XBSJ000102.4--245850 & AGN1 & 0.433 & 2.12$^{+0.13}_{-0.12}$ & $<$0.02 & 0.67 & 43.74 & 40/40 & 47.7 & BSS & Y \\
XBSJ000532.7+200716$^{p}$ & AGN1$^{e}$ & 0.119 & 3.42$^{+0.15}_{-0.15}$ & $<$0.01 & 0.21 & 42.03 & 64/37 & 0.4 & BSS & N,5 \\
XBSJ001831.6+162925 & AGN1 & 0.553 & 2.39$^{+0.06}_{-0.06}$ & $<$0.01 & 0.76 & 44.11 & 123/126 & 54.8 & BSS & Y \\
XBSJ002618.5+105019 & AGN1 & 0.473 & 2.04$^{+0.06}_{-0.06}$ & $<$0.009 & 2.78 & 44.44 & 115/102 & 17.6 & HBSS,BSS & Y \\
XBSJ002637.4+165953 & AGN1 & 0.554 & 2.15$^{+0.07}_{-0.05}$ & $<$0.02 & 1.23 & 44.27 & 123/138 & 81.9 & BSS & Y \\
\hline
\end{longtable}
\begin{list}{}{}
\item Columns: (1) Source name; (2) Class; (3) Redshift; (4) Photon
  index; (5) Intrinsic column density; (6) Observed flux in the 2-10
  keV band, de-absorbed by our Galaxy; (7) Intrinsic luminosity in the
  2-10 keV band; (8) $\chi^{2}$ to number of degrees of freedom; (9)
  Null hypothesis probability; (10) Sample the source belongs to; (11)
  Whether the simple power law is considered as our best fit (Y) or
  not (N). In the case it is not, a number indicates the corresponding
  table's number where our considered best fit is reported.
\item $^{p}$: Null hypothesis probability $<$ 10\%.$^{e}$: Elusive
  AGN. $^{f}$: Fixed parameter. AGN class and redshift in bold face
  mark new optical identifications. Note: Errors and upper limits are at 90\% confidence level. Fluxes and luminosities refer to the MOS2 calibration.
\end{list}
\end{landscape}
}
\end{document}